\documentclass[aps,prb,twocolumn,floatfix,eqsecnum,superscriptaddress,longbibliography]{revtex4-1}

\usepackage{amsfonts}
\usepackage{amsmath}
\usepackage{amssymb}
\usepackage{amsthm}
\usepackage{cases}
\usepackage{url}
\usepackage{framed}
\usepackage{cancel,soul}
\usepackage{graphicx}
\usepackage{multirow}
\usepackage{bm}
\usepackage{float}
\usepackage[shortlabels]{enumitem}
\usepackage[normalem]{ulem}
\usepackage[colorlinks=true,linktoc=page,linkcolor=blue,citecolor=blue]{hyperref}

\newcommand{\bea}{\begin{eqnarray}}
\newcommand{\eea}{\end{eqnarray}}
\def\:={\,\raisebox{0.85pt}{.}\hspace{-2.78pt}\raisebox{2.85pt}{.}\!\!=\,}
\def\=:{\,=\!\!\raisebox{0.85pt}{.}\hspace{-2.78pt}\raisebox{2.85pt}{.}\,}

\begin{document}

\title{Model of spin liquids with and without time-reversal symmetry}

\author{Jyong-Hao Chen}
\affiliation{Condensed Matter Theory Group, Paul Scherrer Institute, 
CH-5232 Villigen PSI, Switzerland}
\author{Christopher Mudry}
\affiliation{Condensed Matter Theory Group, Paul Scherrer Institute, 
CH-5232 Villigen PSI, Switzerland}
\author{Claudio Chamon}
\affiliation{Department of Physics, Boston University, 
Boston, MA, 02215, USA}
\author{A. M. Tsvelik}
\affiliation{
Condensed Matter Physics and Materials Science Division,
Brookhaven National Laboratory, Upton, NY 11973-5000, USA
            }

\date{\today}

\begin{abstract} 
  We study a model in (2+1)-dimensional spacetime that is realized by
  an array of chains, each of which realizes relativistic Majorana fields in 
  (1+1)-dimensional spacetime, coupled via current-current
  interactions. The model is shown to have a lattice realization in an
  array of two-leg quantum spin-$1/2$ ladders. We study the model both
  in the presence and absence of time-reversal symmetry, within a
  mean-field approximation. We find regimes in coupling space where
  Abelian and non-Abelian spin liquid phases are stable. In the case
  when the Hamiltonian is time-reversal symmetric, we find regimes
  where gapped Abelian and non-Abelian chiral phases appear as a
  result of spontaneous breaking of time-reversal symmetry. These
  gapped phases are separated by a discontinuous phase transition. More
  interestingly, we find a regime where a {\it non-chiral gapless
  non-Abelian} spin liquid is stable. The excitations in this regime
  are described by relativistic Majorana fields in (2+1)-dimensional spacetime,
  much as those appearing in the Kitaev honeycomb model, but here emerging in a model of
  coupled spin ladders that does not break $SU(2)$ spin-rotation
  symmetry.
\end{abstract}

\maketitle
\tableofcontents

\section{Motivation and summary of results}
\label{eq: introduction}

The Kalmeyer-Laughlin chiral spin liquid \cite{Kalmeyer87} was the
first example of a connection between the physics of the fractional
quantum Hall (FQH) effect and that of frustrated magnets that do not
order via the spontaneous breaking of a symmetry. Such chiral spin
liquids present exotic features, such as ground state degeneracy on
the torus -- a defining attribute of topological
order~\cite{Wen91a}. The Kitaev honeycomb model~\cite{Kitaev06a}
presents another example of a chiral spin liquid when a gap is opened
by the addition of a magnetic field. The Kitaev model displays, in a
regime of parameters, non-Abelian topological order, where the
quasiparticles obey non-Abelian braiding statistics, as in the
Moore-Read FQH states.~ \cite{Moore91}

Recently, coupled-wire constructions pioneered by Kane and
collaborators~\cite{Mukhopadhyay01,Kane02,Teo14,Kane17,Kane18} have
provided a different approach to the construction of topological
ordered states, in particular both Abelian and non-Abelian FQH
states. These constructions allow one to utilize the powerful
machinery of (1+1)-dimensional conformal field theory (CFT) to describe the
individual quantum wires, which are then coupled to their neighbors to
gap the bulk degrees of freedom of the resulting two-dimensional
system. Gapless chiral modes, described by chiral CFTs~\cite{Wen90a},
are rather naturally obtained in these coupled-wire constructions.

Most of the focus of coupled-wire constructions has been on electronic
systems with a quantized (charge) Hall response. However, one may also
use, instead of quantum wires, quantum spin chains or ladders, which
can also be described by CFTs in their gapless
limits.~\cite{Gorohovsky15,Huang16a,Lecheminant17,Huang17,Chen17,Pereira18} The
result of these coupled-chain (or coupled-ladder) constructions are
gapped chiral spin liquid in (2+1)-dimensional spacetime \cite{Kalmeyer87,Wen89},
much as the electronic wire constructions lead to gapped FQH states.


\begin{figure*}[t]
  \begin{center}
  (a)
  \includegraphics[width=0.53\textwidth]{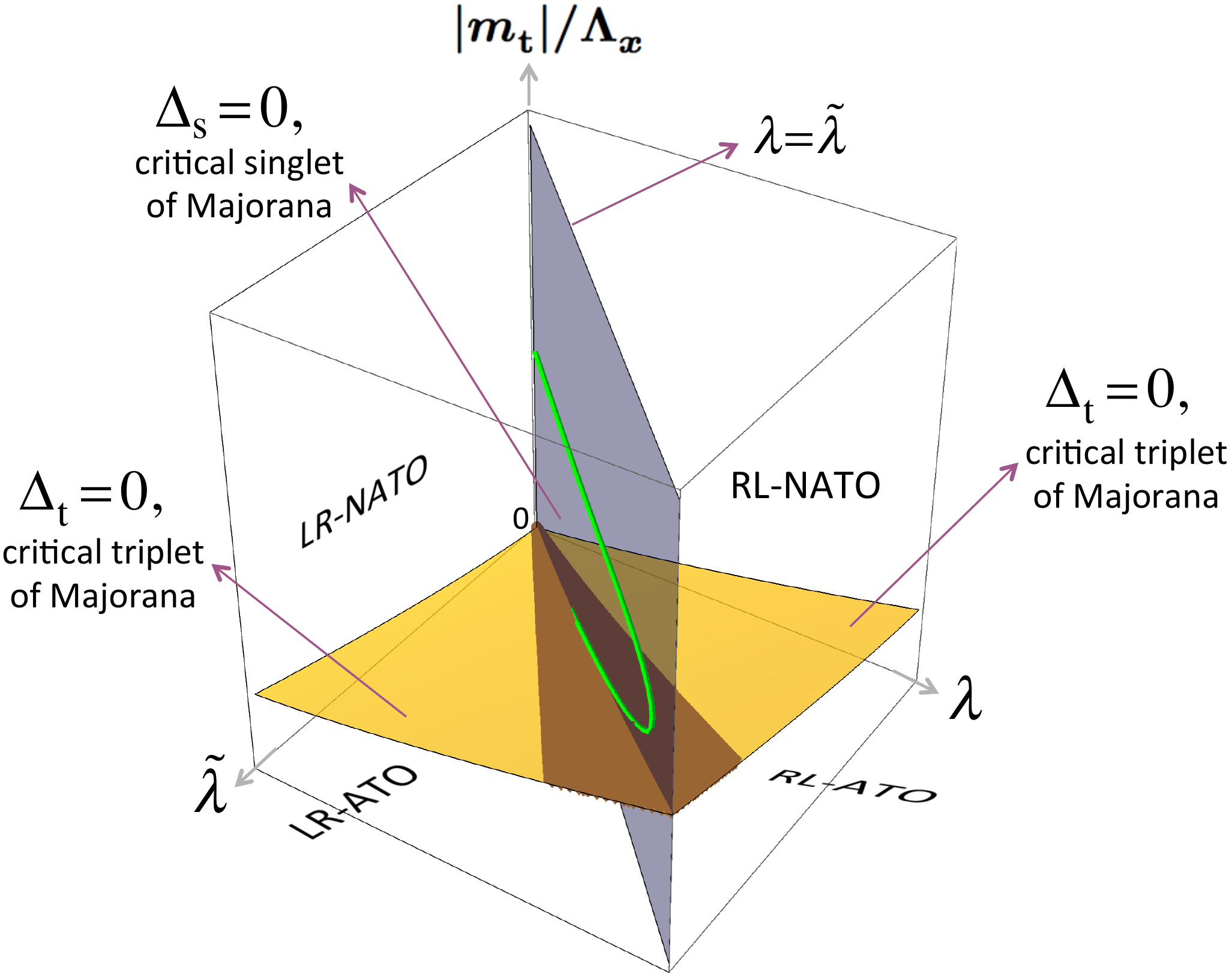}
  (b)
  \includegraphics[width=0.4\textwidth]{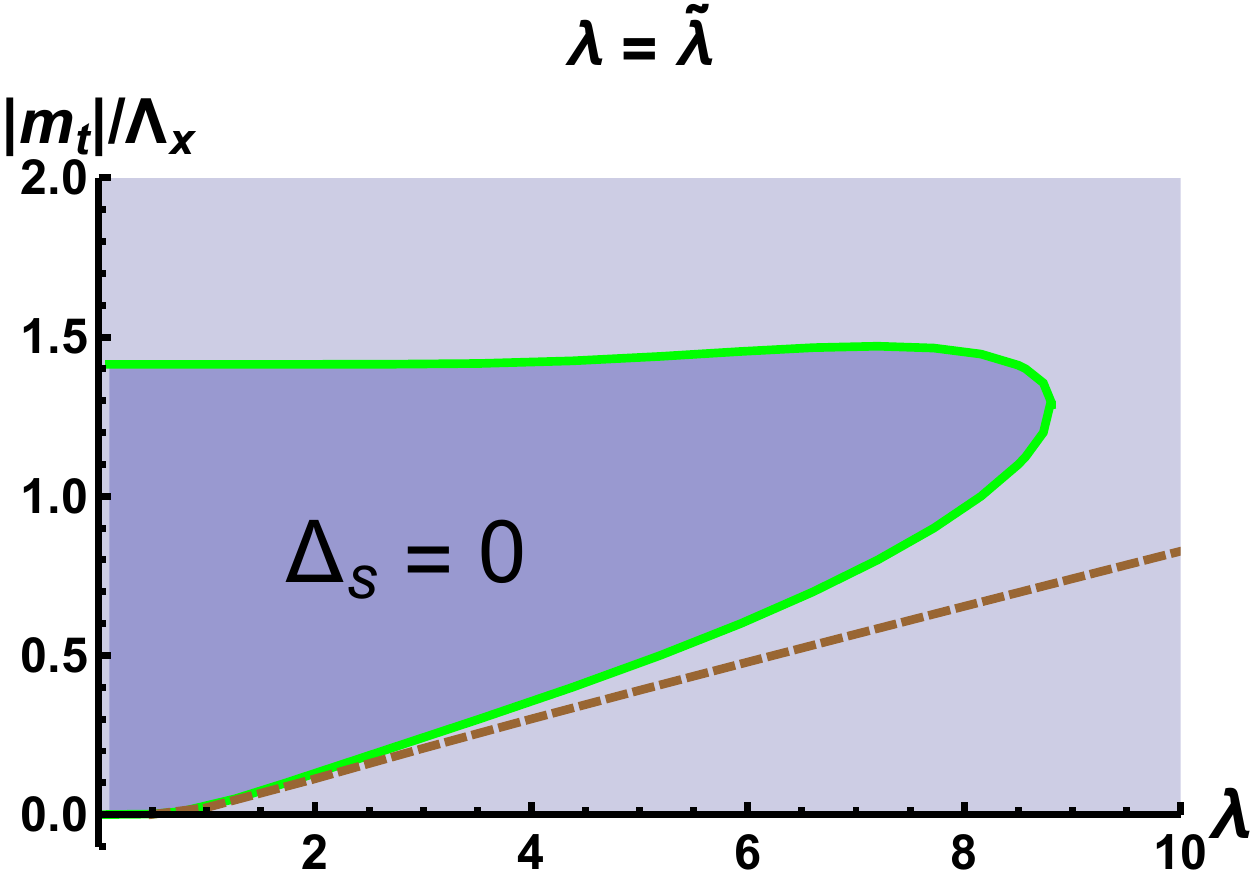}
\caption{(Color online)
(a)
Mean-field phase diagram as a function of 
the three couplings
$\lambda\geq0$, $\widetilde{\lambda}\geq0,$ and
$|m^{\,}_{\mathrm{t}}|/\Lambda^{\,}_{x}$
for the theory defined in Eq.\
(\ref{eq: coupled Majorana theory}) under the assumptions 
(\ref{eq: singlet and triplet anisotropy assumption})
[$\Lambda^{\,}_{x}$ is a momentum cutoff that is introduced
in Eq.\ (\ref{eq: define fermion effective action b})].
The yellow (brown) surface represents those points in coupling space
at which a continuous (discontinuous) mean-field transition separates
two distinct gapped phases of matter,
the mean-field snapshots of an Abelian topological order (ATO) phase
and a non-Abelian topological order (NATO) phase, respectively.
The quadrant $\lambda=\widetilde{\lambda}$
is colored in grey and is explained in the panel (b).
(b)
For $|m^{\,}_{\mathrm{t}}/\Lambda^{\,}_{x}|\geq0$ and $\lambda=\widetilde{\lambda}\geq0$,
the region bounded by the vertical axis and the continuous line
(colored in green) supports
a time-reversal symmetric mean-field solution ($\Phi^{\,}_{-}=0$) 
as the minima of the mean-field potential,
with the vanishing singlet gap $\Delta^{\,}_{\mathrm{s}}=0$
and the non-vanishing triplet gap
$\Delta^{\,}_{\mathrm{t}}=2|m^{\,}_{\mathrm{t}}|\neq0$
defined by Eq.\
(\ref{eq: singlet and triplet gap function}).
Outside of this region, time-reversal symmetry is
spontaneously broken at the mean-field level ($\Phi^{\,}_{-}\neq0$)
with non-vanishing singlet ($\Delta^{\,}_{\mathrm{s}}\neq0$)
and triplet ($\Delta^{\,}_{\mathrm{t}}\neq0$)
gaps defined by Eq.\
(\ref{eq: singlet and triplet gap function}).
The dashed line (colored in brown) is a line of
discontinuous phase transitions
by which $|\Phi^{\,}_{-}|<|m^{\,}_{\mathrm{t}}|$
above the dashed line while $|\Phi^{\,}_{-}|>|m^{\,}_{\mathrm{t}}|$
below the dashed line. It separates 
the mean-field snapshot of an ATO from a NATO phase.
}
\label{Fig: schematic phase diagram}
\end{center}
\end{figure*}


Within this coupled-ladder approach, we presented a model
in Ref.~\onlinecite{Chen17}
that we argued displayed both Abelian and
non-Abelian chiral spin liquid phases. In that model, each two-legged
ladder (central charge $c=2$) can be described using four flavors of
Majorana fields ($c=1/2$ each). These four fields in each ``wire'' can
be split into triplet and singlet representation of $SU(2)$. The Majorana fields
are local within each ``wire'', as quadratic terms (such as
back-scattering) are only allowed {\it inside} the one-dimensional
channels; we denote such mass terms for the triplet and singlet
Majorana fields by $m^{\,}_{\mathrm{t}}$ and $m^{\,}_{\mathrm{s}}$. In
contrast, local inter-ladder interactions are necessarily quartic in
the Majorana fields, and characterized by couplings constants
$\lambda, \widetilde{\lambda}$, which are introduced in
Sec.\ \ref{sec: The Majorana field theory}.
In Ref.~\onlinecite{Chen17} we studied the case $\lambda\ne 0,
\widetilde{\lambda}=0$, which maximally breaks time-reversal symmetry (TRS).  This limit was
analyzed using mean-field theory and a random phase approximation that
started from an exactly solvable limit. Within these approximations,
we obtained the phase diagram of the coupled-ladder system, with its
gapped Abelian and non-Abelian chiral phases.

\subsection*{Summary of results: the phase diagram}

In the present work, we consider generic inter-ladder interactions,
$\lambda, \widetilde{\lambda}\ne 0$, that encompass the case $\lambda=
\widetilde{\lambda}$ where TRS is not explicitly broken, and we search
for {\it non-chiral} spin liquids within the coupled ladder system. We
find a rather rich phase diagram within a mean-field approximation,
which we depict in Fig.~\ref{Fig: schematic phase diagram}. The phase
diagram includes the gapped chiral Abelian and non-Abelian phases when
TRS is either explicitly or spontaneously broken. In the case where
TRS is spontaneously broken, we find that the Abelian and non-Abelian
phases are separated by a discontinuous transition. More interestingly,
we identify a regime where TRS remains unbroken, leading to a gapless
non-chiral spin liquid. At the gapless region, the Majorana fields
acquire dispersion in the direction perpendicular to the ladders,
yielding a pair of 2D Majorana cones. We thus find an example of a
spin system with full $SU(2)$ spin-rotation invariance that supports
a non-chiral spin liquid phase with gapless Majoranas
as in the phase~B of Kitaev honeycomb model. However, $SU(2)$
spin-rotation symmetry is absent in the Kitaev honeycomb model.

The paper is organized as follows. We present
the model of coupled Majorana fields
and analyze several of its symmetries in
Sec.\ \ref{sec: The Majorana field theory}.
We then study the model within a mean-field treatment in
Sec.\ \ref{sec: The mean-field approach}.
In Sec.\ \ref{sec: Implications for lattice models},
we discuss possible implications of the mean-field phase diagram
for a model of coupled spin-ladder, whose couplings are contained
within the coupled Majorana field theory.
We summarize our results
in Sec.\ \ref{sec: Summary}.


\section{Model of coupled Majorana field theories}
\label{sec: The Majorana field theory}

\subsection{Definition}

Our quantum-field theory is built from four species
(labeled by $\mu=0,\cdots,3$) of Majorana fields
whose support is $(1+1)$-dimensional spacetime.
We will call this building-block a ``ladder''.
This terminology is justified by the fact that we find
in Sec.\ \ref{sec: Implications for lattice models}
a spin-1/2 ladder that regularizes this quantum field theory.
We then consider $n$ independent copies of the
Majorana quantum-field theory
in $(1+1)$-dimensional spacetime with the kinetic
Hamiltonian density
\begin{align}
&
\widehat{\mathcal{H}}^{\,}_{0}\:=
\sum^{n}_{\mathtt{m}=1}
\sum^{3}_{\mu=0}
\frac{\mathrm{i}}{2}
v^{\,}_{\mu}
\left(
\widehat{\chi}^{\mu}_{\mathrm{L},\mathtt{m}}
\partial^{\,}_{x}
\widehat{\chi}^{\mu}_{\mathrm{L},\mathtt{m}}
-
\widehat{\chi}^{\mu}_{\mathrm{R},\mathtt{m}}
\partial^{\,}_{x}
\widehat{\chi}^{\mu}_{\mathrm{R},\mathtt{m}}
\right),
\label{eq: coupled Majorana theory kinetic}
\end{align}
where the velocities $v^{\,}_{\mu}$ are real valued
and
$\mathrm{L},\mathrm{R}$
denotes the left- and right-movers, respectively.
The Majorana fields, $(\widehat{\chi}^{\mu}_{\mathrm{M},\mathtt{m}})^{*}=\widehat{\chi}^{\mu}_{\mathrm{M},\mathtt{m}}$,
obey the equal-time anti-commutators
$
\left\{
\widehat{\chi}^{\mu}_{\mathrm{M},\mathtt{m}}(x),
\widehat{\chi}^{\mu'}_{\mathrm{M}',\mathtt{m}'}(x')
\right\}=
\delta^{\,}_{\mathrm{M}\mathrm{M}'}\,
\delta^{\,}_{\mathtt{m}\mathtt{m}'}\,	
\delta^{\,}_{\mu\mu'}\,
\delta(x-x'),
$
with
$\mu,\mu'=0,\cdots3,$
$\mathrm{M},\mathrm{M}'=\mathrm{L},\mathrm{R}$,  
$\mathtt{m},\mathtt{m}'=1,\cdots,n$,
and $0\leq x \leq L^{\,}_{x}.$

Besides the kinetic term (\ref{eq: coupled Majorana theory kinetic}), 
we assume that there is 
a back-scattering term 
with real valued couplings $m^{\,}_{\mu}$ ($\mu=0,\cdots,3$) inside each ladder
\begin{align}
&
\widehat{\mathcal{H}}^{\,}_{\mathrm{intra}-\mathrm{ladder}}\:=
\sum^{n}_{\mathtt{m}=1}
\sum^{3}_{\mu=0}
\mathrm{i}\,m^{\,}_{\mu}\,
\widehat{\chi}^{\mu}_{\mathrm{L},\mathtt{m}}
\widehat{\chi}^{\mu}_{\mathrm{R},\mathtt{m}}.
\label{eq: coupled Majorana theory intraladder}
\end{align}
We then couple consecutive ladders by
considering inter-ladder quartic interactions with real valued
coupling constants $\lambda$ and $\widetilde{\lambda}$
\begin{subequations}
\label{eq: coupled Majorana theory interladder}
\begin{align}
&
\widehat{\mathcal{H}}^{\,}_{\mathrm{inter}-\mathrm{ladder}}\:=
\sum^{n-1}_{\mathtt{m}=1}
\left(
\widehat{\mathcal{H}}^{\,}_{\lambda,\mathtt{m}}
+
\widehat{\mathcal{H}}^{\,}_{\widetilde{\lambda},\mathtt{m}}
\right),
\label{eq: coupled Majorana theory inter-ladder a}
\\
&
\widehat{\mathcal{H}}^{\,}_{\lambda,\mathtt{m}}\:=
\frac{\lambda\vphantom{\widetilde{\lambda}}}{4}
\left(
\sum^{3}_{\mu=0}
\widehat{\chi}^{\mu}_{\mathrm{L},\mathtt{m}}\,
\widehat{\chi}^{\mu}_{\mathrm{R},\mathtt{m}+1}\,
\right)^{2},
\label{eq: coupled Majorana theory inter-ladder b}
\\
&
\widehat{\mathcal{H}}^{\,}_{\widetilde{\lambda},\mathtt{m}}\:=
\frac{\widetilde{\lambda}}{4}
\left(
\sum^{3}_{\mu=0}
\widehat{\chi}^{\mu}_{\mathrm{R},\mathtt{m}}\,
\widehat{\chi}^{\mu}_{\mathrm{L},\mathtt{m}+1}\,
\right)^{2}.
\label{eq: coupled Majorana theory inter-ladder c}
\end{align}
\end{subequations}
Each $\widehat{\mathcal{H}}^{\,}_{\lambda,\mathtt{m}}$ 
and
$\widehat{\mathcal{H}}^{\,}_{\widetilde{\lambda},\mathtt{m}}$ 
term alone
is the $O(4)$ Gross-Neveu-like quartic interaction.

The final
Hamiltonian density is
\begin{align}
&
\widehat{\mathcal{H}}\:=
\widehat{\mathcal{H}}^{\,}_{0}
+
\widehat{\mathcal{H}}^{\,}_{\mathrm{intra}-\mathrm{ladder}}
+
\widehat{\mathcal{H}}^{\,}_{\mathrm{inter}-\mathrm{ladder}},
\label{eq: coupled Majorana theory}
\end{align}
with 
$\widehat{\mathcal{H}}^{\,}_{0}$,
$\widehat{\mathcal{H}}^{\,}_{\mathrm{intra}-\mathrm{ladder}}$
and
$\widehat{\mathcal{H}}^{\,}_{\mathrm{inter}-\mathrm{ladder}}$
defined in Eq.\ 
(\ref{eq: coupled Majorana theory kinetic}),
(\ref{eq: coupled Majorana theory intraladder})
and
(\ref{eq: coupled Majorana theory interladder}),
respectively. 

The limit $\widetilde{\lambda}=0$ in the Hamiltonian density
(\ref{eq: coupled Majorana theory})
was considered in Ref.\ \onlinecite{Chen17}. 
This regime corresponds (with the singlet mass
$m^{\,}_{\mathrm{s}}=0$) to the planar region
  $(\lambda\geq0,\widetilde{\lambda}=0,m^{\,}_{\mathrm{t}}>0,m^{\,}_{\mathrm{s}}=0)$
  in Fig.~\ref{Fig: schematic phase diagram},
  where ATO and NATO are the abbreviations
for ``Abelian topological order''
and ``non-Abelian topological order'', respectively.
A telltale to distinguish these phases is the
central charge $c$ of edge
states: For Abelian phases, $c$ is
necessarily integer; instead, if $c$ is fractional, the phase is
necessarily non-Abelian. (Notice that it is possible to have integer
$c$ for non-Abelian phases, for instance direct sums of models with
fractional $c$'s that add up to an integer.) In our model,
the signatures of these
phases at the mean-field level are the following.
The edge states of a mean-field snapshot of the ATO phase
are quadruplet
of right-moving (left-moving) Majorana fermions
$\widehat{\chi}^{\mu}_{\mathrm{R},1}$ ($\widehat{\chi}^{\mu}_{\mathrm{L},n}$)
on the first (last) edge for $\mu=0,1,2,3$, yielding
$c=4\times 1/2=2\in\mathbb{Z}$.
The edge states of a mean-field snapshot of the NATO phase
consist of the singlet Majorana modes $\widehat{\chi}^{0}_{\mathrm{R},1}$
and $\widehat{\chi}^{0}_{\mathrm{L},n}$, with $c=1\times 1/2=1/2\notin\mathbb{Z}$.

The goal of this work is to study the generic case where both
$\lambda$ and $\widetilde{\lambda}$ are {\it non-zero}. The phase
  diagram in Fig.~\ref{Fig: schematic phase diagram}
  is mirror symmetric about the plane
$\lambda=\widetilde{\lambda}$, and we shall be particularly interested
in the limit $\lambda=\widetilde{\lambda}$ at which the Hamiltonian
density (\ref{eq: coupled Majorana theory}) is invariant under TRS.

\subsection{Symmetries}

Reversal of time is implemented by the $\mathtt{m}$-resolved
\textit{antiunitary} $\mathbb{Z}^{\,}_{2}$ transformation by which
\begin{equation}
\begin{split}
&
\widehat{\chi}^{\mu}_{\mathrm{L},\mathtt{m}}(x)\;\mapsto\;
\widehat{\chi}^{\mu}_{\mathrm{R},\mathtt{m}}(x),
\\&
\widehat{\chi}^{\mu}_{\mathrm{R},\mathtt{m}}(x)\;\mapsto\;
\widehat{\chi}^{\mu}_{\mathrm{L},\mathtt{m}}(x),
\\&
\mathrm{i}\;\mapsto\;-\mathrm{i},
\end{split}
\label{eq: trs symmetries}
\end{equation}
for any $\mu=0,\cdots,3$,
$\mathtt{m}=1,\cdots,n$,
and
$0\leq x\leq L^{\,}_{x}$.

The Hamiltonian density (\ref{eq: coupled Majorana theory})
has more symmetries.
First, for arbitrary values of the masses and the couplings,
the Hamiltonian density (\ref{eq: coupled Majorana theory})
is invariant under
\begin{equation}
\widehat{\chi}^{\mu}_{\mathrm{M},\mathtt{m}}(x)\;\mapsto\;
\sigma^{\mu}\,
\widehat{\chi}^{\mu}_{\mathrm{M},\mathtt{m}}(x),
\qquad
\sigma^{\mu}=\pm1,
\label{eq: symmetries a}
\end{equation}
for any $\mu=0,\cdots,3$,
$\mathrm{M}=\mathrm{L},\mathrm{R}$,
$\mathtt{m}=1,\cdots,n$,
and
$0\leq x\leq L^{\,}_{x}$.
Second, it is also invariant under the
$\mathtt{m}$-resolved (local)
$\mathbb{Z}^{\,}_{2}$
transformation by which
\begin{equation}
\widehat{\chi}^{\mu}_{\mathrm{M},\mathtt{m}}(x)\;\mapsto\;
\sigma^{\,}_{\mathtt{m}}\,
\widehat{\chi}^{\mu}_{\mathrm{M},\mathtt{m}}(x),
\qquad
\sigma^{\,}_{\mathtt{m}}=\pm1,
\label{eq: symmetries b}
\end{equation}
for any $\mu=0,\cdots,3$,
$\mathrm{M}=\mathrm{L},\mathrm{R}$,
$\mathtt{m}=1,\cdots,n$,
and
$0\leq x\leq L^{\,}_{x}$.

Whenever the underlying lattice regularization
of the Hamiltonian density
(\ref{eq: coupled Majorana theory})
is endowed with a global $SU(2)$ symmetry,
we will impose the conditions
\begin{equation}
\begin{split}
&
v^{\,}_{0}\equiv v^{\,}_{\mathrm{s}}\equiv v,\qquad
m^{\,}_{0}\equiv m^{\,}_{\mathrm{s}}=0,
\\
&
v^{\,}_{a}\equiv v^{\,}_{\mathrm{t}}\equiv v,\qquad
m^{\,}_{a}\equiv m^{\,}_{\mathrm{t}},\qquad
a=1,2,3,
\label{eq: singlet and triplet anisotropy assumption}
\end{split}
\end{equation}
where $\mathrm{s}$ and $\mathrm{t}$ stands for ``singlet'' and ``triplet'',
respectively.

\section{Mean-field approach}
\label{sec: The mean-field approach}

\subsection{Two auxiliary scalar fields}
\label{subsec: Hubbard-Stratonovich transformation}

We will treat the inter-ladder quartic interactions
(\ref{eq: coupled Majorana theory interladder})
by performing a Hubbard-Stratonovich transformation.
To this end,  we employ the Euclidean path-integral formalism and
introduce two real-valued auxiliary scalar fields, 
$\phi^{\,}_{\mathtt{m},\mathtt{m}+1}$ and $\tilde{\phi}^{\,}_{\mathtt{m},\mathtt{m}+1}$
for $\mathtt{m}=1,\cdots,n-1.$
The model (\ref{eq: coupled Majorana theory}) can then be written as
\begin{subequations}
\label{eq: def partition fct with HS decoupling}
\begin{align}
&
Z\:=
\int\mathcal{D}[\phi,\tilde{\phi}]
\int\mathcal{D}[\chi^{0},\chi^{1},\chi^{2},\chi^{3}]\,
e^{-S},
\label{eq: def partition fct with HS decoupling a}
\\
&
S\:=
\int\limits_{0}^{\beta}\mathrm{d}\tau\,
\int\limits_{0}^{L^{\,}_{x}}\mathrm{d}x\,
\sum_{\mathtt{m}=1}^{L^{\,}_{y}/\mathfrak{a}^{\,}_{y}}
\left(
\mathcal{L}^{\,}_{\mathrm{f},\mathtt{m}}
+
\mathcal{L}^{\,}_{\mathrm{b},\mathtt{m}}
+
\mathcal{L}^{\,}_{\mathrm{fb},\mathtt{m}}
\right),
\label{eq: def partition fct with HS decoupling b}
\\
&
\mathcal{L}^{\,}_{\mathrm{f},\mathtt{m}}\:=
\frac{1}{2}
\sum^{3}_{\mu=0}
\Big[
\chi^{\mu}_{\mathrm{L},\mathtt{m}}
\left(\partial^{\,}_{\tau}+\mathrm{i}v^{\,}_{\mu}\partial^{\,}_{x}\right)
\chi^{\mu}_{\mathrm{L},\mathtt{m}}
\nonumber\\
&\hspace{6em}
+
\chi^{\mu}_{\mathrm{R},\mathtt{m}}
\left(\partial^{\,}_{\tau}-\mathrm{i}v^{\,}_{\mu}\partial^{\,}_{x}\right)
\chi^{\mu}_{\mathrm{R},\mathtt{m}}
\Big]
\nonumber\\
&\hspace{3em}
+
\sum_{\mu=0}^{3}
\mathrm{i}\,
m^{\,}_{\mu}
\chi^{\mu}_{\mathrm{L},\mathtt{m}}\,
\chi^{\mu}_{\mathrm{R},\mathtt{m}},
\label{eq: def partition fct with HS decoupling c}
\\
&
\mathcal{L}^{\,}_{\mathrm{b},\mathtt{m}}\:=
\frac{1}{4\,\lambda}(\phi^{\,}_{\mathtt{m},\mathtt{m}+1})^{2}
+
\frac{1}{4\,\widetilde{\lambda}}(\tilde{\phi}^{\,}_{\mathtt{m},\mathtt{m}+1})^{2},
\label{eq: def partition fct with HS decoupling d}
\\
&
\mathcal{L}^{\,}_{\mathrm{fb},\mathtt{m}}\:=
\sum_{\mu=0}^{3}
\frac{1}{2}
\left(
-\mathrm{i}\chi^{\mu}_{\mathrm{L},\mathtt{m}}\,\chi^{\mu}_{\mathrm{R},\mathtt{m}+1}
\right)
\phi^{\,}_{\mathtt{m},\mathtt{m}+1}
\nonumber\\
&
\hphantom{
\mathcal{L}^{\,}_{\chi,\phi,\mathtt{m}}\:=
         }
+
\sum_{\mu=0}^{3}
\frac{1}{2}
\left(
-\mathrm{i}\chi^{\mu}_{\mathrm{R},\mathtt{m}}\,\chi^{\mu}_{\mathrm{L},\mathtt{m}+1}
\right)
\tilde{\phi}^{\,}_{\mathtt{m},\mathtt{m}+1}.
\label{eq: def partition fct with HS decoupling e}
\end{align}
\end{subequations}
Here, $\beta$ is the inverse temperature and
$\mathfrak{a}^{\,}_{y}$ is the spacing
between two consecutive ladders.

\subsection{Symmetries}

The action
(\ref{eq: def partition fct with HS decoupling b})
with $\lambda=\widetilde{\lambda}$
is invariant under the $\mathtt{m}$-resolved antiunitary time-reversal
transformation [c.f.\ Eq.\ (\ref{eq: trs symmetries})]
\begin{equation}
\begin{split}
  \chi^{\mu}_{\mathrm{M},\mathtt{m}}(\tau,x)
  &\;\mapsto\;
  \chi^{\mu}_{\mathrm{M}',\mathtt{m}}(\tau,x),
  \qquad
  \mathrm{M}\neq\mathrm{M}',
  \\
  \phi^{\,}_{\mathtt{m},\mathtt{m}+1}(\tau,x)
  &\;\mapsto\;
  -\,\tilde{\phi}^{\,}_{\mathtt{m},\mathtt{m}+1}(\tau,x),
  \\
  \tilde{\phi}^{\,}_{\mathtt{m},\mathtt{m}+1}(\tau,x)
  &\;\mapsto\;
  -\,\phi^{\,}_{\mathtt{m},\mathtt{m}+1}(\tau,x),
\end{split}
\label{eq: symmetries Majorana Grassmann c}
\end{equation}
~for any $\mu=0,\cdots,3$,
$\mathrm{M}=\mathrm{L},\mathrm{R}$,
$\mathtt{m}=1,\cdots,n$,
$0\leq\tau\leq\beta$,
and
$0\leq x\leq L^{\,}_{x}$.

The action
(\ref{eq: def partition fct with HS decoupling b})
has the following additional symmetries. First, the $\mu$-resolved Majorana
parity is conserved owing to the
symmetry of the action $S$
(\ref{eq: def partition fct with HS decoupling b})
under the $\mathbb{Z}^{\,}_{2}$ transformation 
[c.f. Eq. (\ref{eq: symmetries a})]
\begin{equation}
\chi^{\mu}_{\mathrm{M},\mathtt{m}}(\tau,x)\;\mapsto\;
\sigma^{\mu}\,
\chi^{\mu}_{\mathrm{M},\mathtt{m}}(\tau,x),
\qquad
\sigma^{\mu}=\pm1,
\label{eq: symmetries Majorana Grassmann a}
\end{equation}
for any $\mu=0,\cdots,3$,
$\mathrm{M}=\mathrm{L},\mathrm{R}$,
$\mathtt{m}=1,\cdots,n$,
$0\leq\tau\leq\beta$,
and
$0\leq x\leq L^{\,}_{x}$.

Second, the action
(\ref{eq: def partition fct with HS decoupling b})
is invariant under the $\mathtt{m}$-resolved $\mathbb{Z}^{\,}_{2}$ transformation
[c.f. Eq. (\ref{eq: symmetries b})]
\begin{equation}
\begin{split}
  \chi^{\mu}_{\mathrm{M},\mathtt{m}}(\tau,x)
  &\;\mapsto\;
  \sigma^{\,}_{\mathtt{m}}\,
  \chi^{\mu}_{\mathrm{M},\mathtt{m}}(\tau,x),
  \qquad
  \sigma^{\,}_{\mathtt{m}}=\pm1,
  \\
  \phi^{\,}_{\mathtt{m},\mathtt{m}+1}(\tau,x)
  &\;\mapsto\;
  \sigma^{\,}_{\mathtt{m}}\,
  \sigma^{\,}_{\mathtt{m}+1}\,
  \phi^{\,}_{\mathtt{m},\mathtt{m}+1}(\tau,x),
  \\
  \tilde{\phi}^{\,}_{\mathtt{m},\mathtt{m}+1}(\tau,x)
  &\;\mapsto\;
  \sigma^{\,}_{\mathtt{m}}\,
  \sigma^{\,}_{\mathtt{m}+1}\,
  \tilde{\phi}^{\,}_{\mathtt{m},\mathtt{m}+1}(\tau,x),
\end{split}
\label{eq: symmetries Majorana Grassmann b}
\end{equation}
for any $\mu=0,\cdots,3$,
$\mathrm{M}=\mathrm{L},\mathrm{R}$,
$\mathtt{m}=1,\cdots,n$,
$0\leq\tau\leq\beta$,
and
$0\leq x\leq L^{\,}_{x}$.

\subsection{Mean-field single-particle Hamiltonian}
\label{subsec: Mean-field single-particle Hamiltonian}

We do the mean-field approximation by which the 
Hubbard-Stratonovich 
fields $\phi$ and $\tilde{\phi}$
are assumed independent of the spacetime coordinates
$(\tau,x)$ and the ladder index $\mathtt{m}$,
\begin{align}
&
\phi^{\,}_{\mathtt{m},\mathtt{m}+1}(\tau,x)\equiv \phi,
\qquad
\tilde{\phi}^{\,}_{\mathtt{m},\mathtt{m}+1}(\tau,x)\equiv \tilde{\phi}.
\label{eq: mean-field assumption}
\end{align}
In what follows, we will ignore sign fluctuations of these Hubbard-Stratonovich fields
$\phi$ and $\tilde{\phi}$,
since as was demonstrated in Ref.\ \onlinecite {Assaad16}, 
where fermions coupled to a $\mathbb{Z}^{\,}_{2}$ gauge field on a square lattice were studied, 
such fluctuations are irrelevant.
If so, the action from (\ref{eq: def partition fct with HS decoupling d})
simplifies to
\begin{align}
S^{\,}_{\mathrm{b}}\:=&
\int\limits_{0}^{\beta}\mathrm{d}\tau
\int\limits_{0}^{L^{\,}_{x}}\mathrm{d}x\,
\sum_{\mathtt{m}=1}^{L^{\,}_{y}/\mathfrak{a}^{\,}_{y}}\,
\mathcal{L}^{\,}_{\mathrm{b},\mathtt{m}}
\nonumber\\
=&\,
\beta\,L^{\,}_{x}\,\frac{L^{\,}_{y}}{\mathfrak{a}^{\,}_{y}}
\left(
\frac{1}{4\lambda}\phi^{2}
+
\frac{1}{4\widetilde{\lambda}}\tilde{\phi}^{2}
\right).
\end{align}

\begin{widetext}
We proceed by imposing periodic boundary condition
along the $y$-direction,
$\chi^{\mu}_{\mathrm{M},n+1}\equiv\chi^{\mu}_{\mathrm{M},1}$
for $\mathrm{M}=\mathrm{L},\mathrm{R}$,
and by performing the Fourier transformation
\begin{subequations}
\begin{align}
S^{\,}_{\mathrm{f}}
+
S^{\,}_{\mathrm{fb}}\:=
\int\limits_{0}^{\beta}\mathrm{d}\tau
\int\limits_{0}^{L^{\,}_{x}}\mathrm{d}x\,
\sum_{\mathtt{m}=1}^{L^{\,}_{y}/\mathfrak{a}^{\,}_{y}}\,
\left(
\mathcal{L}^{\,}_{\mathrm{f},\mathtt{m}}
+
\mathcal{L}^{\,}_{\mathrm{fb},\mathtt{m}}
\right)
=
\sum^{\,}_{\omega,\bm{k}}\, 
\sum^{3}_{\mu=0}
\frac{1}{2}
\left(\chi^{\mu}_{-\omega,-\bm{k}}\right)^{\mathrm{T}}
\left(
\mathrm{i}\omega\sigma^{\,}_{0}
+
\widehat{H}^{\mathrm{MF}}_{\mu,\bm{k}}
\right)
\chi^{\mu}_{\omega,\bm{k}},
\end{align}
where
$
\chi^{\mu}_{\omega,\bm{k}}:=
\left(
\chi^{\mu}_{\mathrm{R},\omega,\bm{k}},
\chi^{\mu}_{\mathrm{L},\omega,\bm{k}}
\right)^{\mathsf{T}}
$
for each flavor $\mu=0,\cdots,3$, 
the mean-field Majorana Hamiltonian is
\begin{align}
\widehat{H}^{\mathrm{MF}}_{\mu,\bm{k}}\:=
-v^{\,}_{\mu}k^{\,}_{x}\,
\sigma^{\,}_{3}
-
\Phi^{\,}_{+}
\sin
\left(
k^{\,}_{y}
\mathfrak{a}^{\,}_{y}
\right)
\sigma^{\,}_{1}
+
\left[
m^{\,}_{\mu}
-
\Phi^{\,}_{-}
\cos
\left(
k^{\,}_{y}
\mathfrak{a}^{\,}_{y}
\right)
\right]\,
\sigma^{\,}_{2},
\label{eq: mean-field Majorana single-particle Hamiltonian}
\end{align}
and we have introduced the linear combinations
\begin{align}
\Phi^{\,}_{\pm}\:=\frac{1}{2}\left(\phi\pm\tilde{\phi}\right),
\label{eq: define plus and minu basis}
\end{align}
\end{subequations}
for the auxiliary scalar fields. Here,
$\sigma^{\,}_{1}$,
$\sigma^{\,}_{2}$,
and $\sigma^{\,}_{3}$ are Pauli matrices, while
$\sigma^{\,}_{0}$ is the $2\times2$ identity matrix. 
\end{widetext}

We can diagonalize the $2\times2$ single-particle Hamiltonian 
(\ref{eq: mean-field Majorana single-particle Hamiltonian})
for each flavor $\mu=0,\cdots,3$.
There follows eight branches of mean-field excitations
with the dispersions (we have set $\mathfrak{a}^{\,}_{y}=1$)
\begin{subequations}
\begin{align}
&\varepsilon^{\,}_{\mu,\pm}(k^{\,}_{x},k^{\,}_{y})\:=
\pm\varepsilon^{\,}_{\mu}(k^{\,}_{x},k^{\,}_{y}),
\\
&\varepsilon^{\,}_{\mu}(k^{\,}_{x},k^{\,}_{y})\:=
\sqrt{
v^{2}_{\mu}k^{2}_{x}
+
\left(
m^{\,}_{\mu}
-
\Phi^{\,}_{-}
\cos k^{\,}_{y}
\right)^{2}
+
\Phi^{2}_{+}
\sin^{2}k^{\,}_{y}
}.
\end{align}
\end{subequations}
We see that the eight branches fall into four pairs of
particle-hole symmetric bands.
For arbitrary value of $k^{\,}_{x}$ and $k^{\,}_{y},$
the mean-field Majorana direct gap is defined by
\begin{align}
\Delta^{\,}_{\mu}(k^{\,}_{x},k^{\,}_{y})\:=&\,
\varepsilon^{\,}_{+,\mu}
(k^{\,}_{x},k^{\,}_{y})
-
\varepsilon^{\,}_{-,\mu}
(k^{\,}_{x},k^{\,}_{y})
\nonumber\\
=&\,
2\,\varepsilon^{\,}_{\mu}
(k^{\,}_{x},k^{\,}_{y}).
\end{align}
In the vicinity of 
$(k^{\,}_{x}=0,k^{\,}_{y}=0)$ 
and
$(k^{\,}_{x}=0,k^{\,}_{y}=\pi)$,
the mean-field Majorana direct gaps are,
\begin{subequations}
\label{eq: gap function}
\begin{align}
\Delta^{\,}_{\mu,}(0,0)
=
2\left|m^{\,}_{\mu}-\Phi^{\,}_{-}\right|
=
2\left|m^{\,}_{\mu}-\frac{\phi}{2}+\frac{\tilde{\phi}}{2}\right|,
\label{eq: gap function a}
\end{align}
and
\begin{align}
\Delta^{\,}_{\mu}(0,\pi)
=
2\left|m^{\,}_{\mu}+\Phi^{\,}_{-}\right|
=
2\left|m^{\,}_{\mu}+\frac{\phi}{2}-\frac{\tilde{\phi}}{2}\right|,
\label{eq: gap function b}
\end{align}
\end{subequations}
respectively.
The miminum of the two gap functions (\ref{eq: gap function})
is
\begin{align}
\Delta^{\,}_{\mu}\:=&\,
2\left||m^{\,}_{\mu}|-\left|\Phi^{\,}_{-}\right|\right|
=
2\left||m^{\,}_{\mu}|-\frac{1}{2}\left|\phi-\tilde{\phi}\right|\right|.
\label{eq: gap function summary}
\end{align}

\subsection{Linearized spectrum}

\begin{widetext}
The physics captured by the 
mean-field Majorana single-particle Hamiltonian 
(\ref{eq: mean-field Majorana single-particle Hamiltonian})
becomes more transparent upon
linearizing the latter
around the gap closing points
$(k^{\,}_{x},k^{\,}_{y})=(0,0)$
and
$(k^{\,}_{x},k^{\,}_{y})=(0,\pi/\mathfrak{a}^{\,}_{y})$,
respectively. One finds
\begin{subequations}
\label{eq: mean-field Majorana single-particle Hamiltonian linearized}
\begin{align}
&
\widehat{H}^{\mathrm{MF}}_{\mu,k^{\,}_{x}=0+p^{\,}_{x},k^{\,}_{y}=0+p^{\,}_{y}}\approx
-
v^{\,}_{\mu}\,
p^{\,}_{x}\,
\sigma^{\,}_{3}
-
\mathfrak{a}^{\,}_{y}\,
\Phi^{\,}_{+}\,
p^{\,}_{y}\,
\sigma^{\,}_{1}
+
\left(
m^{\,}_{\mu}
-
\Phi^{\,}_{-}
\right)\,
\sigma^{\,}_{2},
\label{eq: mean-field Majorana single-particle Hamiltonian linearized a}
\\
&
\widehat{H}^{\mathrm{MF}}_
{\mu,k^{\,}_{x}=0+p^{\,}_{x},k^{\,}_{y}=(\pi/\mathfrak{a}^{\,}_{y})+p^{\,}_{y}}
\approx
-
v^{\,}_{\mu}\,
p^{\,}_{x}\,
\sigma^{\,}_{3}
+
\mathfrak{a}^{\,}_{y}\,
\Phi^{\,}_{+}\,
p^{\,}_{y}\,
\sigma^{\,}_{1}
+
\left(
m^{\,}_{\mu}
+
\Phi^{\,}_{-}
\right)\,
\sigma^{\,}_{2}.
\label{eq: mean-field Majorana single-particle Hamiltonian linearized b}
\end{align}
\end{subequations}
Accordingly,
$\mathfrak{a}^{\,}_{y}\Phi^{\,}_{+}$
plays the role of the Fermi velocity in the $y$-direction.
Furthermore, we find that the single-particle Majorana gap is
2$\left|m^{\,}_{\mu}\mp\Phi^{\,}_{-}\right|$,
in agreement
with Eqs.\
(\ref{eq: gap function a})
and
(\ref{eq: gap function b}).
We now combine these linearized
mean-field Majorana single-particle Hamiltonian 
into the $4\times4$ matrix
\begin{subequations}
\begin{align}
\widehat{H}^{\mathrm{MF},\mathrm{lin}}_{\mu,\bm{p}}\:=&\,
\begin{pmatrix}
\widehat{H}^{\mathrm{MF}}_{\mu,k^{\,}_{x}=0+p^{\,}_{x},k^{\,}_{y}=0+p^{\,}_{y}}
&
0^{\,}_{2\times2}
\\
0^{\,}_{2\times2}
&
\widehat{H}^{\mathrm{MF}}_
{\mu,k^{\,}_{x}=0+p^{\,}_{x},k^{\,}_{y}=(\pi/\mathfrak{a}^{\,}_{y})+p^{\,}_{y}} 
\end{pmatrix}
\nonumber\\
=&\,
-
v^{\,}_{x,\mu}\,
p^{\,}_{x}\,
\sigma^{\,}_{3}
\otimes
\tau^{\,}_{0}
-
v^{\,}_{y}\,
p^{\,}_{y}\,
\sigma^{\,}_{1}
\otimes
\tau^{\,}_{3}
+
m^{\,}_{\mu}\,
\sigma^{\,}_{2}
\otimes
\tau^{\,}_{0}
-
\Phi^{\,}_{-}\,
\sigma^{\,}_{2}
\otimes
\tau^{\,}_{3},
\label{eq: linearized 4 by 4 hamiltonian}
\end{align}
where we have defined
\begin{align}
v^{\,}_{x,\mu}\:=v^{\,}_{\mu},
\qquad
v^{\,}_{y}\:=
\mathfrak{a}^{\,}_{y}
\Phi^{\,}_{+}.
\label{eq: define velocity in y direction}
\end{align}
\end{subequations}
\end{widetext}
This is an anisotropic single-particle Dirac Hamiltonian.
The anisotropy enters through the two distinct Fermi velocities
(\ref{eq: define velocity in y direction}),
with the velocity along the $y$ direction emerging from the
non-vanishing value $\Phi^{\,}_{+}$ for the bonding
linear combination of the Hubbard-Stratonovich fields.
There are two competing masses,
$m^{\,}_{\mu}$ and the anti-bonding linear combination $\Phi^{\,}_{-}$
of the Hubbard-Stratonovich fields
that measures the amount by which the mean-field breaks
time-reversal symmetry. These masses compete because they
multiply two $4\times4$ matrices that commute,
\begin{equation}
\left[
\sigma^{\,}_{2}\otimes\tau^{\,}_{0},
\sigma^{\,}_{2}\otimes\tau^{\,}_{3}
\right]=0.
\end{equation}
The mass term
$m^{\,}_{\mu}\,\sigma^{\,}_{2}\otimes\tau^{\,}_{0}$
breaks a unitary $\mathbb{Z}^{\,}_{2}$ symmetry represented by conjugation with
\begin{equation}
\hat{\mathcal{I}}=\sigma^{\,}_{3}\otimes\tau^{\,}_{1}.
\end{equation}
The mass term $\Phi^{\,}_{-}\,\sigma^{\,}_{2}\otimes\tau^{\,}_{3}$
breaks time-reversal symmetry that is represented by
conjugation with
\begin{equation}
\hat{\mathcal{T}}=\sigma^{\,}_{1}\otimes\tau^{\,}_{1}\,\mathsf{K},
\end{equation}
where $\mathsf{K}$ denotes the complex conjugation.

The competition between the
mass terms
$m^{\,}_{\mu}\,\sigma^{\,}_{2}\otimes\tau^{\,}_{0}$
and
$\Phi^{\,}_{-}\,\sigma^{\,}_{2}\otimes\tau^{\,}_{3}$
implies a gap closing (i.e., continuous) transition when
\begin{equation}
|m^{\,}_{\mu}|=|\Phi^{\,}_{-}|
\end{equation}
that separates two single-particle insulating phases.
As shown by Haldane \cite{Haldane88}, the Chern numbers for the pair of band
resolved by the flavor index $\mu$ is $\pm1$ when
\begin{equation}
|m^{\,}_{\mu}|<|\Phi^{\,}_{-}|.
\end{equation}
This single-particle insulating phase
realizes a Chern insulator at half-filling.
When open boundary conditions are imposed, channel $\mu$ contributes
one (Majorana) chiral edge state. The Chern numbers for the pair of band
resolved by the flavor index $\mu$ have vanishing Chern numbers when
\begin{equation}
|m^{\,}_{\mu}|>|\Phi^{\,}_{-}|.
\end{equation}
This single-particle insulating phase is topologically trivial
at half-filling. 
Gapless boundary states are not generic when
open boundary conditions are imposed.



\subsection{Mean-field potential}
\label{subsec: Integrate out Majorana fields and the effective potential}

After integrating out the Majorana fields
and expressing the scalar fields $\phi$ and $\tilde{\phi}$
in terms of $\Phi^{\,}_{\pm}$ by using Eq.\
(\ref{eq: define plus and minu basis}),
the partition function
(\ref{eq: def partition fct with HS decoupling})
becomes
$
Z\propto
\int\mathcal{D}[\Phi^{\,}_{+},\Phi^{\,}_{-}]\,
e^{-S^{\,}_{\mathrm{eff}}},
$
where 
\begin{subequations}
\label{eq: define effective action}
\begin{align}
&
S^{\,}_{\mathrm{eff}}\:=
S^{\,}_{\mathrm{B}}
+
S^{\,}_{\mathrm{F}},
\label{eq: define effective action a}
\\
&
S^{\,}_{\mathrm{B}}\:=
\frac{\beta\,L^{\,}_{x}\,L^{\,}_{y}}{\mathfrak{a}^{\,}_{y}}
\left[
\frac{1}{4\lambda\vphantom{\widetilde{\lambda}}}
\left(
\Phi^{\,}_{+}
+
\Phi^{\,}_{-}
\right)^{2}
+
\frac{1}{4\widetilde{\lambda}}
\left(
\Phi^{\,}_{+}
-
\Phi^{\,}_{-}
\right)^{2}
\right],
\label{eq: define effective action b}
\\
&
S^{\,}_{\mathrm{F}}\:=
-
\frac{1}{2}
\sum^{3}_{\mu=0}
\sum^{\,}_{\omega,\bm{k}}
\nonumber\\
&
\log
\left(
-
\omega^{2}
-
v^{2}_{\mu}k^{2}_{x}
-
\left(m^{\,}_{\mu}-\Phi^{\,}_{-}\cos q\right)^{2}
-
\Phi^{2}_{+}\sin^{2} q
\right),
\label{eq: define effective action c}
\end{align}
with $q\:=k^{\,}_{y}\mathfrak{a}^{\,}_{y}.$
\end{subequations}

When $\lambda=\widetilde{\lambda}$,
the action (\ref{eq: define effective action}) is invariant under a
global antiunitary $\mathbb{Z}^{\,}_{2}$
transformation defined by
\begin{equation}
\Phi^{\,}_{+}\;\mapsto\;\,-\Phi^{\,}_{+},
\qquad
\Phi^{\,}_{-}\;\mapsto\;\,\Phi^{\,}_{-},
\qquad
\mathrm{i}\;\mapsto\;\,-\mathrm{i}.
\label{eq: trs symmetries big phi basis}
\end{equation}
This transformation is the mean-field counterpart to
the time-reversal transformation defined in 
(\ref{eq: symmetries Majorana Grassmann c}).
We note that the $\mu$-resolved global Majorana parity represented by 
the $\mathbb{Z}^{\,}_{2}$ transformation
(\ref{eq: symmetries Majorana Grassmann a}) 
is invisible in the action (\ref{eq: define effective action})
as we have integrated out Majorana fields.
The $\mathtt{m}$-resolved $\mathbb{Z}^{\,}_{2}$
transformation (\ref{eq: symmetries Majorana Grassmann b})
is also invisible in the action (\ref{eq: define effective action})
since $\sigma^{\,}_{\mathtt{m}}=\sigma^{\,}_{\mathtt{m}+1}$
for any $\mathtt{m}=1,\cdots,n$ 
under the mean-field approximation (\ref{eq: mean-field assumption}).

We are interested in the zero temperature and thermodynamic limit 
$\beta\to\infty$, $L^{\,}_{x}\to\infty$, and $L^{\,}_{y}\to\infty$
of the effective action (\ref{eq: define effective action}).
The summations then become integrals in three-dimensional spacetime. 

The Bosonic contribution to the mean-field
potential is defined by
\begin{align}
V^{\,}_{\mathrm{MF},\mathrm{B}}\:=&\,
\frac{\mathfrak{a}^{\,}_{y}}{\beta\,L^{\,}_{x}\,L^{\,}_{y}}
S^{\,}_{B}
\nonumber\\
=&\,
\frac{1}{4\lambda\vphantom{\widetilde{\lambda}}}
\left(
\Phi^{\,}_{+}
+
\Phi^{\,}_{-}
\right)^{2}
+
\frac{1}{4\widetilde{\lambda}}
\left(
\Phi^{\,}_{+}
-
\Phi^{\,}_{-}
\right)^{2},
\label{eq: define boson effective action}
\end{align}
where $S^{\,}_{B}$ is given by Eq.\ (\ref{eq: define effective action b}).

Similarly, the Fermionic contribution to the
mean-field
potential is
\begin{subequations}
\label{eq: define fermion effective action}
\begin{align}
&
V^{\,}_{\mathrm{MF},\mathrm{F}}\:=
\frac{\mathfrak{a}^{\,}_{y}}{\beta\,L^{\,}_{x}\,L^{\,}_{y}}
S^{\,}_{\mathrm{F}}\:=
\sum^{3}_{\mu=0}
V^{\mu}_{\mathrm{eff},\mathrm{F}},
\label{eq: define fermion effective action a}
\end{align}
where 
$S^{\,}_{\mathrm{F}}$ is given by Eq.\ (\ref{eq: define effective action c})
and we have defined
\begin{align}
&
V^{\mu}_{\mathrm{eff},\mathrm{F}}\:=
-
\frac{1}{2}
\int\limits^{+\infty}_{-\infty}\frac{\mathrm{d}\omega}{2\pi}
\int\limits^{+\Lambda^{\,}_{x}}_{-\Lambda^{\,}_{x}}\frac{\mathrm{d}k^{\,}_{x}}{2\pi}
\int\limits^{+\pi}_{-\pi}\frac{\mathrm{d}q}{2\pi}
\times
\nonumber\\
&\,\qquad
\log
\left(
\omega^{2}
+
v^{2}_{\mu}k^{2}_{x}
+
\left(m^{\,}_{\mu}-\Phi^{\,}_{-}\cos q\right)^{2}
+
\Phi^{2}_{+}\sin^{2} q
\right),
\label{eq: define fermion effective action b}
\end{align}
with $\Lambda^{\,}_{x}$ a momentum cutoff.
\end{subequations}
After performing the integrals over the Matsubara frequency $\omega$ 
and over the momentum $k^{\,}_{x}$,%
~\footnote{
In Eq.\ (\ref{eq: define fermion effective action b}),
the integral over $\omega$
can be carried out by using the following formula for definite integral
\cite{Stone00}
$$
\int\limits^{\infty}_{-\infty}\frac{\mathrm{d}\omega}{2\pi}
\ln\left(\omega^2+k^{2}+A^2\right)=
\sqrt{k^{2}+A^{2}}
+
\mathrm{const},
$$
where the constant is formally infinity.
          }
we are left with
\begin{widetext}
\begin{subequations}
\label{eq: fermion effective action final}
\begin{align}
&
V^{\mu}_{\mathrm{eff},\mathrm{F}}=
-
\Lambda^{2}_{x}\,
\frac{1}{2\pi}\,
\frac{1}{v^{\,}_{\mu}}\,
\int\limits^{+\pi}_{-\pi}
\frac{\mathrm{d}q}{2\pi}\times
\left[
\frac{1}{4}\,
\sqrt{F^{2}_{\mu}(q)
+
\frac{1}{4}}
+
\frac{1}{2}\,
F^{2}_{\mu}(q)\,
\ln\left(\frac{1}{2}+\sqrt{F^{2}_{\mu}(q)+\frac{1}{4}}\right)
-
\frac{1}{2}\,
F^{2}_{\mu}(q)\,
\ln|F^{\,}_{\mu}(q)|
\right],
\label{eq: fermion effective action final a}
\end{align}
where 
\begin{align}
F^{\,}_{\mu}(q)\:=
\sqrt{
\left(
\frac{m^{\,}_{\mu}}{\Lambda^{\,}_{x}}
-
\frac{\Phi^{\,}_{-}}{\Lambda^{\,}_{x}}\cos q
\right)^{2}
+
\left(
\frac{\Phi^{\,}_{+}}{\Lambda^{\,}_{x}}
\right)^{2}
\sin^{2} q	
     }.
\label{eq: fermion effective action final b}
\end{align}
\end{subequations}
\end{widetext}

Finally, the total mean-field potential
$V^{\,}_{\mathrm{MF}}$
is the addition of the bosonic mean-field potential
$V^{\,}_{\mathrm{MF},\mathrm{B}}$ 
(\ref{eq: define boson effective action})
to the fermionic mean-field potential
$V^{\,}_{\mathrm{MF},\mathrm{F}}$ 
(\ref{eq: define fermion effective action}), i.e.,
\begin{align}
V^{\,}_{\mathrm{MF}}\:=&\,
V^{\,}_{\mathrm{MF},\mathrm{B}}+V^{\,}_{\mathrm{MF},\mathrm{F}}.
\label{eq: define total effective action}
\end{align}
It is more convenient to 
rewrite $V^{\,}_{\mathrm{MF}}$
(\ref{eq: define total effective action})
into the dimensionless form
\begin{widetext}
\begin{align}
v^{\,}_{\mathrm{MF}}(\Phi^{\,}_{+},\Phi^{\,}_{-})\:=&
\Lambda^{-2}_{x}\times
V^{\,}_{\mathrm{MF}}(\Phi^{\,}_{+},\Phi^{\,}_{-})
\nonumber\\
=&
\frac{1}{4\lambda\vphantom{\widetilde{\lambda}}}
\left(
\frac{\Phi^{\,}_{+}}{\Lambda^{\,}_{x}}
+
\frac{\Phi^{\,}_{-}}{\Lambda^{\,}_{x}}
\right)^{2}
+
\frac{1}{4\widetilde{\lambda}}
\left(
\frac{\Phi^{\,}_{+}}{\Lambda^{\,}_{x}}
-
\frac{\Phi^{\,}_{-}}{\Lambda^{\,}_{x}}
\right)^{2}
\nonumber\\
&
-\frac{1}{2\pi}
\sum^{3}_{\mu=0}
\frac{1}{v^{\,}_{\mu}}
\int\limits^{+\pi}_{-\pi}
\frac{\mathrm{d}q}{2\pi}
\times
\left[
\frac{1}{4}\sqrt{F^{2}_{\mu}(q)+\frac{1}{4}}
+
\frac{1}{2}F^{2}_{\mu}(q)\,
\ln\left(\frac{1}{2}+\sqrt{F^{2}_{\mu}(q)+\frac{1}{4}}\right)
-
\frac{1}{2}F^{2}_{\mu}(q)\,\ln|F^{\,}_{\mu}(q)|
\right],
\label{eq: define total effective action dimensionless}
\end{align}
\end{widetext}
with $F^{\,}_{\mu}(q)$
defined in Eq.\ (\ref{eq: fermion effective action final b}).

We observe  that the mean-field potential
(\ref{eq: define total effective action dimensionless})
is invariant under 
\begin{align}
m^{\,}_{\mu}\to-m^{\,}_{\mu},
\quad
\Phi^{\,}_{+}\to-\Phi^{\,}_{+},
\quad
\Phi^{\,}_{-}\to-\Phi^{\,}_{-}.
\end{align}
Thus, without loss of generality, we shall assume that
$m^{\,}_{\mu},\Phi^{\,}_{+}\geq0$, while $\Phi^{\,}_{-}\in\mathbb{R}$.

\subsection{Saddle-point equations}
\label{subsec: The saddle-point equations}

\begin{widetext}
The saddle-point equations stem from the first-order derivative of $V^{\,}_{\mathrm{eff}}$ 
(\ref{eq: define total effective action dimensionless}).
After performing in closed form
the integrals over the Matsubara frequency $\omega$ and over $k^{\,}_{x}$,
the saddle-point equations are
\begin{subequations}
\label{eq: the saddle-point equation ms neq 0 befor integral new basis}
\begin{align}
0=&\, 
\left(
\frac{1}{2\lambda\vphantom{\widetilde{\lambda}}}
+
\frac{1}{2\widetilde{\lambda}}
\right)
\Phi^{\,}_{+}
+
\left(
\frac{1}{2\lambda\vphantom{\widetilde{\lambda}}}
-
\frac{1}{2\widetilde{\lambda}}
\right)
\Phi^{\,}_{-}
-
\frac{1}{2\pi}
\sum^{3}_{\mu=0}
\left[
\frac{1}{v^{\,}_{\mu}}
\int\limits_{-\pi}^{+\pi}\frac{\mathrm{d}q}{2\pi}
\Phi^{\,}_{+}\,\sin^{2}q
\times
\mathrm{arcsinh}
\left(
\frac{v^{\,}_{\mu}}{2F^{\,}_{\mu}(q)}
    \right)
\right],
\label{eq: the saddle-point equation ms neq 0 befor integral new basis a}
\\
0=&\, 
\left(
\frac{1}{2\lambda\vphantom{\widetilde{\lambda}}}
-
\frac{1}{2\widetilde{\lambda}}
\right)
\Phi^{\,}_{+}
+
\left(
\frac{1}{2\lambda\vphantom{\widetilde{\lambda}}}
+
\frac{1}{2\widetilde{\lambda}}
\right)
\Phi^{\,}_{-}
-
\frac{1}{2\pi}
\sum^{3}_{\mu=0}
\left[
\frac{1}{v^{\,}_{\mu}}
\int\limits_{-\pi}^{+\pi}\frac{\mathrm{d}q}{2\pi}
\left(
\Phi^{\,}_{-}\cos q
-m^{\,}_{\mu}
\right)\times\cos q
\times
\mathrm{arcsinh}
\left(
\frac{v^{\,}_{\mu}}{2F^{\,}_{\mu}(q)}
    \right)
\right],
\label{eq: the saddle-point equation ms neq 0 befor integral new basis b}
\end{align}
where $F^{\,}_{\mu}(q)$
is defined in Eq.\ (\ref{eq: fermion effective action final b}).
\end{subequations}
\end{widetext}

For simplicity, we assume a hidden $SU(2)$ symmetry that implies
that the conditions (\ref{eq: singlet and triplet anisotropy assumption})
must hold (see Sec.\ \ref{sec: Implications for lattice models}).
For simplicity,
$v^{\,}_{\mathrm{s}}=v^{\,}_{\mathrm{t}}\equiv v\equiv1$.
We also assume that $m^{\,}_{\mathrm{s}}=0$,
a consequence of fine tuning at a quantum critical point
of a microscopic building block of the model
(see Sec.\ \ref{Numerical study of a two-leg ladder}).
We solve for $(\Phi^{\,}_{+},\Phi^{\,}_{-})$
in Eq.\
(\ref{eq: the saddle-point equation ms neq 0 befor integral new basis})
numerically for arbitrary value of
$\lambda$, $\widetilde{\lambda}$,
and $\frac{m^{\,}_{\mathrm{t}}}{\Lambda^{\,}_{x}}$.
As we are only interested in local minima of the saddle-point equations
(\ref{eq: the saddle-point equation ms neq 0 befor integral new basis}),
we use the Hessian matrix
\begin{align}
\mathbb{H}^{\,}_{\mathrm{Hess}}\:=
\begin{pmatrix}
\frac{\partial^{2}V^{\,}_{\mathrm{MF}}}{\partial\Phi^{2}_{+}}
&
\frac{\partial^{2}V^{\,}_{\mathrm{MF}}}
    {\partial\Phi^{\,}_{+}\partial\Phi^{\,}_{-}}
\\
\frac{\partial^{2}V^{\,}_{\mathrm{MF}}}
     {\partial\Phi^{\,}_{-}\partial\Phi^{\,}_{+}}
&
\frac{\partial^{2}V^{\,}_{\mathrm{MF}}}{\partial\Phi^{2}_{-}}
\end{pmatrix}
\end{align}
and demand that it is positive definite. A solution
$(\Phi^{\,}_{+},\Phi^{\,}_{-})$
of Eq.\
(\ref{eq: the saddle-point equation ms neq 0 befor integral new basis})
is stable if the Hessian matrix evaluated at
$(\Phi^{\,}_{+},\Phi^{\,}_{-})$ is positive definite.

\begin{figure*}[t]
\begin{center}
(a)
\includegraphics[width=0.405\textwidth]{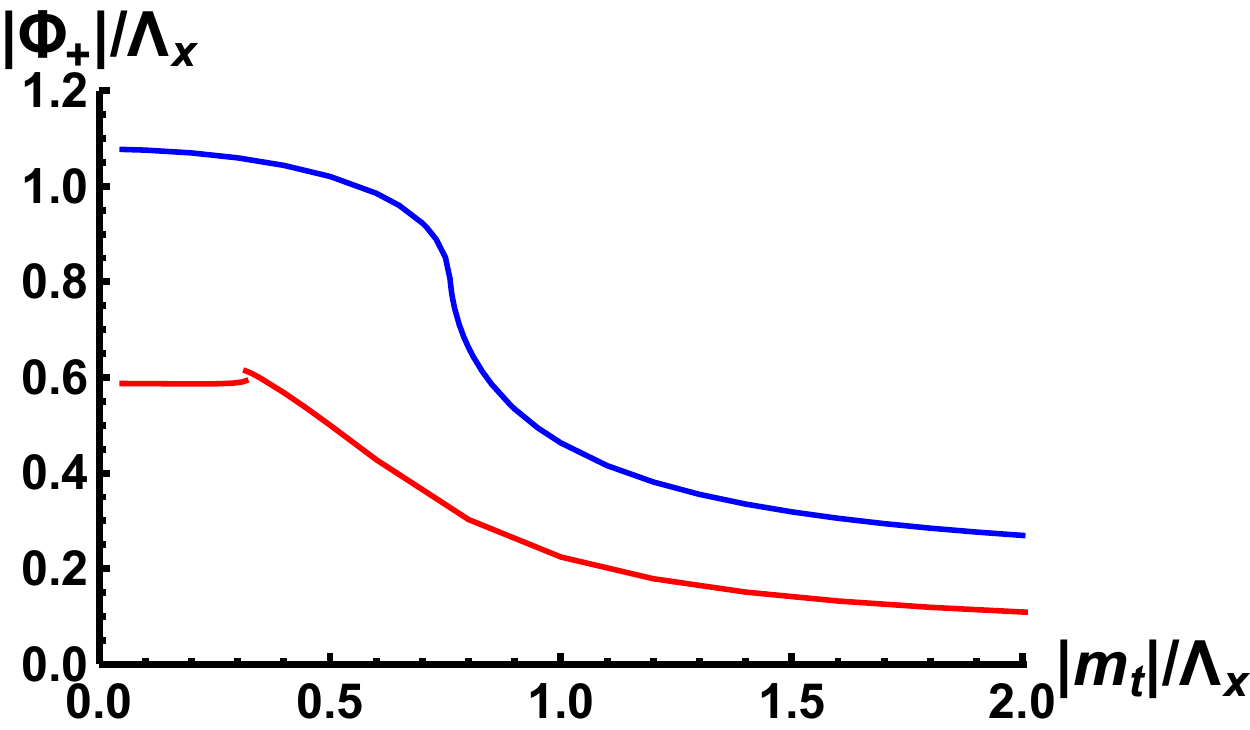}
(b)
\includegraphics[width=0.405\textwidth]{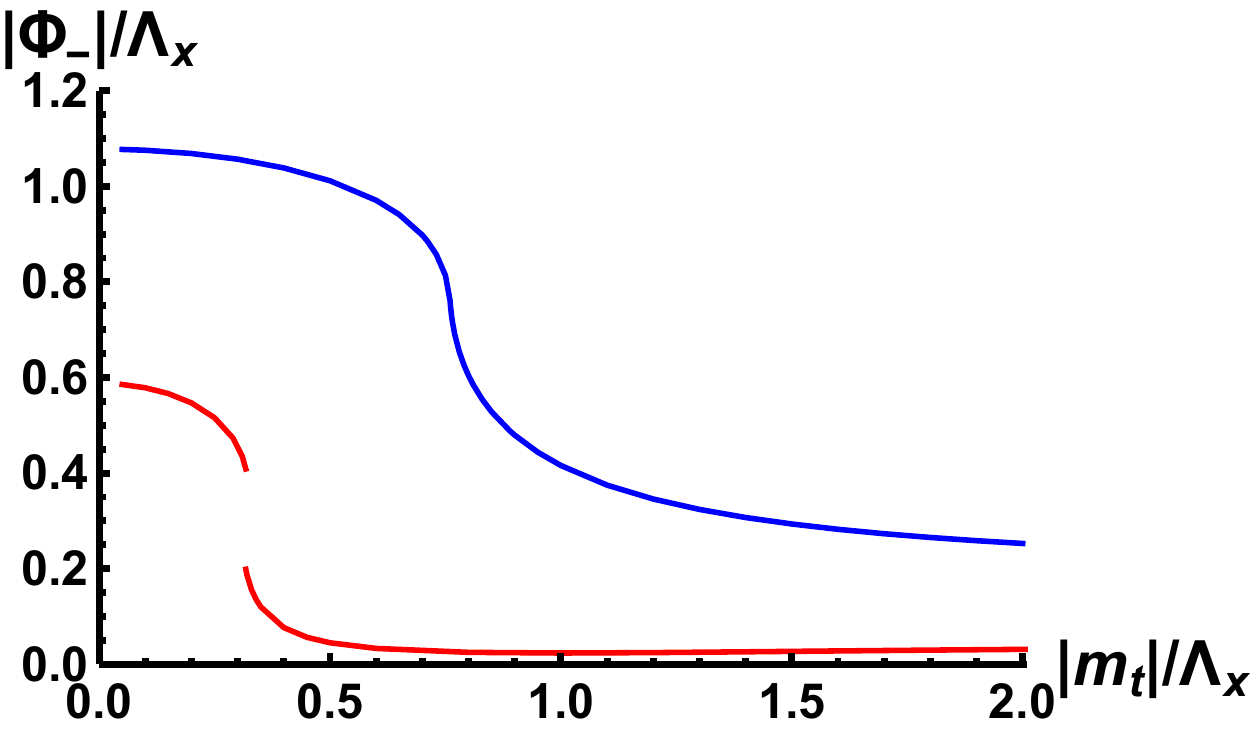}
\\
(c)
\includegraphics[width=0.41\textwidth]{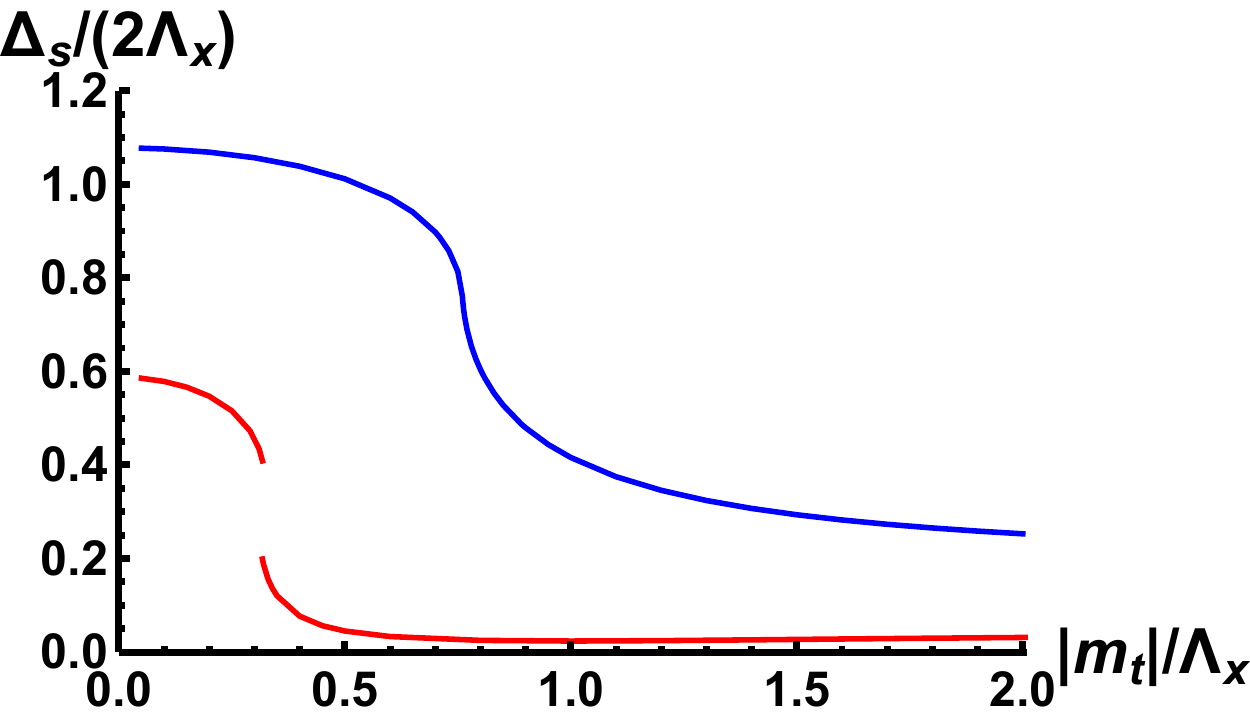}
(d)
\includegraphics[width=0.41\textwidth]{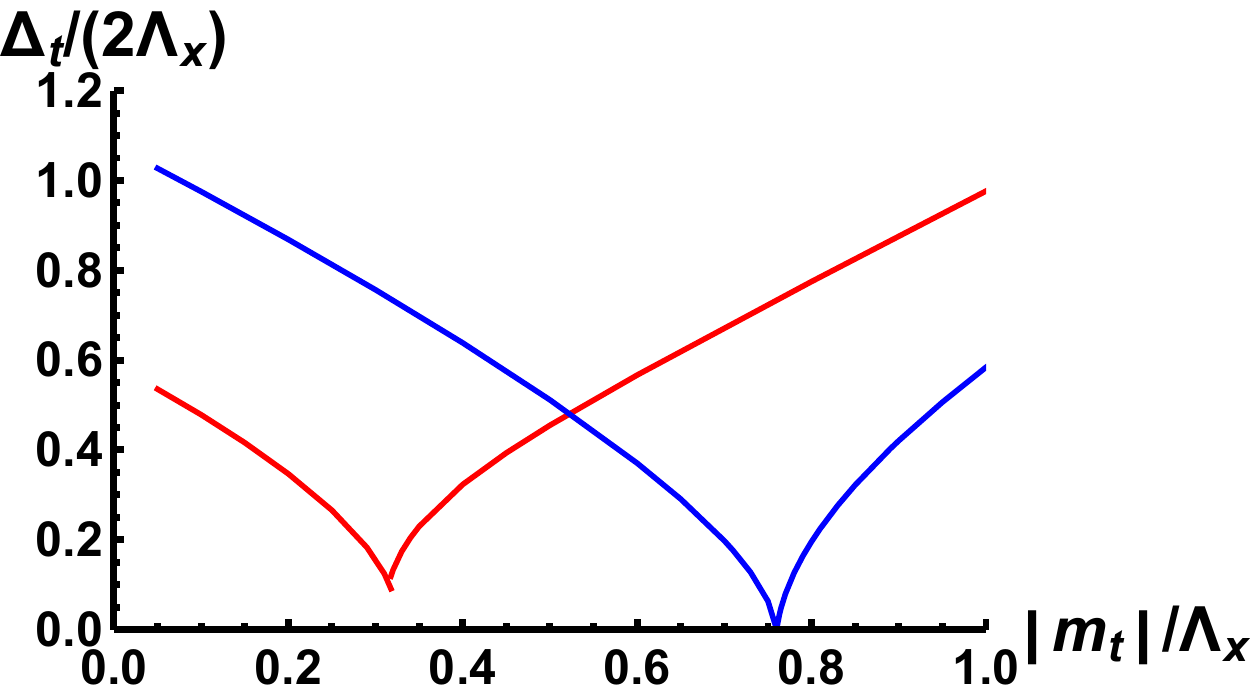}
\caption{(Color online)
Blue curve:
we fix
$(\lambda, \widetilde{\lambda})=(7,1)$ 
in Fig.\ \ref{Fig: schematic phase diagram}
so as to strongly break time-reversal symmetry.
Red~curve:
we fix
$(\lambda,\widetilde{\lambda})=4\sqrt{2}(\sin\theta,\cos\theta)$
with $\theta=23\pi/90$ 
in Fig.\ \ref{Fig: schematic phase diagram}
so as to weakly break time-reversal symmetry.
(a)-(b)
Continuous and discontinuous 
dependence on $m^{\,}_{\mathrm{t}}$
of the stable solution $\Phi^{\,}_{+}$ and $\Phi^{\,}_{-}$ 
to the saddle point equations 
(\ref{eq: the saddle-point equation ms neq 0 befor integral new basis}).
(c)-(d)
Continuous and discontinuous 
dependence on $m^{\,}_{\mathrm{t}}$
of the singlet gap $\Delta^{\,}_{\mathrm{s}}$ and
the triplet gap $\Delta^{\,}_{\mathrm{t}}$ 
given by Eq.\ (\ref{eq: singlet and triplet gap function}).
The triplet gap $\Delta^{\,}_{\mathrm{t}}$ represented by the blue curve vanishes 
at $|m^{\,}_{\mathrm{t}}|/\Lambda^{\,}_{x}\approx0.76$ 
that signals a continuous quantum phase transition.
\label{Fig: details for transition between ATO and NATO}
         }
\end{center}
\end{figure*}

\begin{figure*}[t]
\begin{center}
(a)
\includegraphics[width=0.41\textwidth]{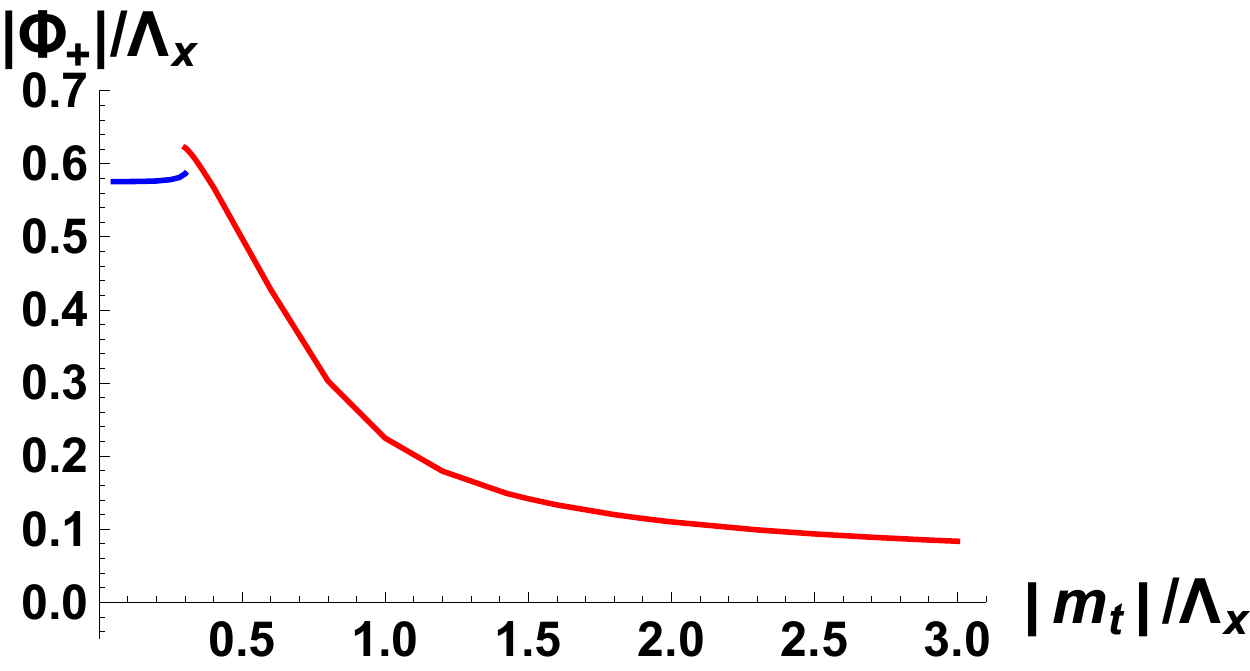}
(b)
\includegraphics[width=0.41\textwidth]{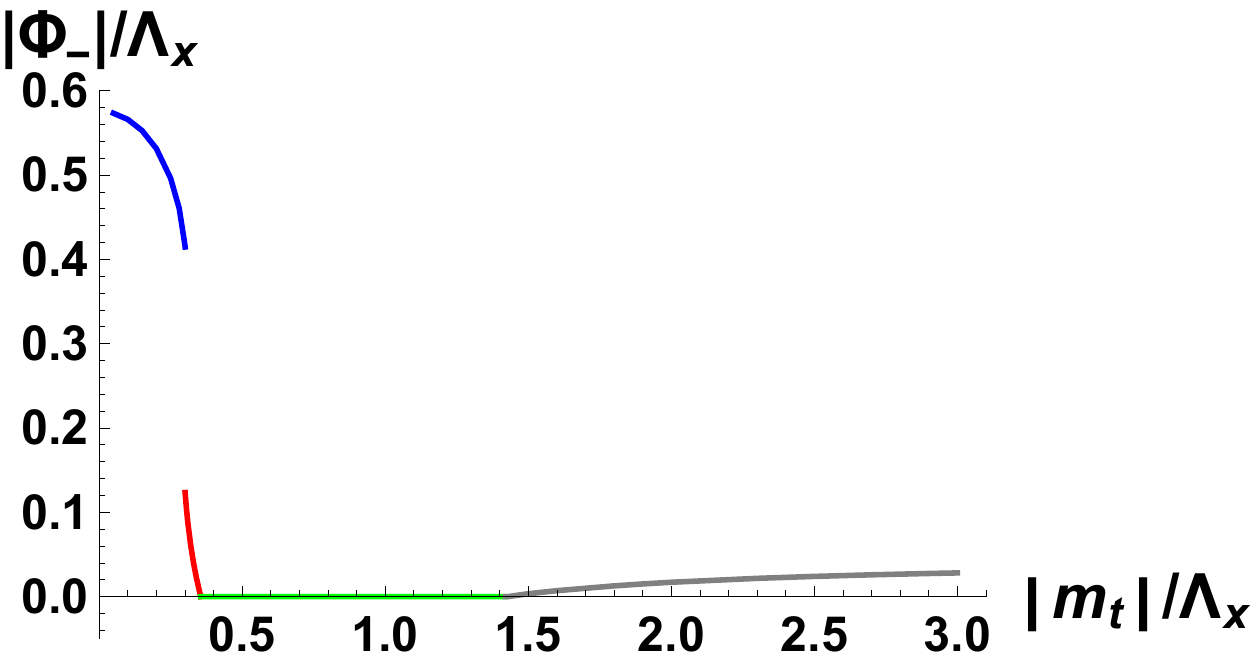}
\\
(c)
\includegraphics[width=0.41\textwidth]{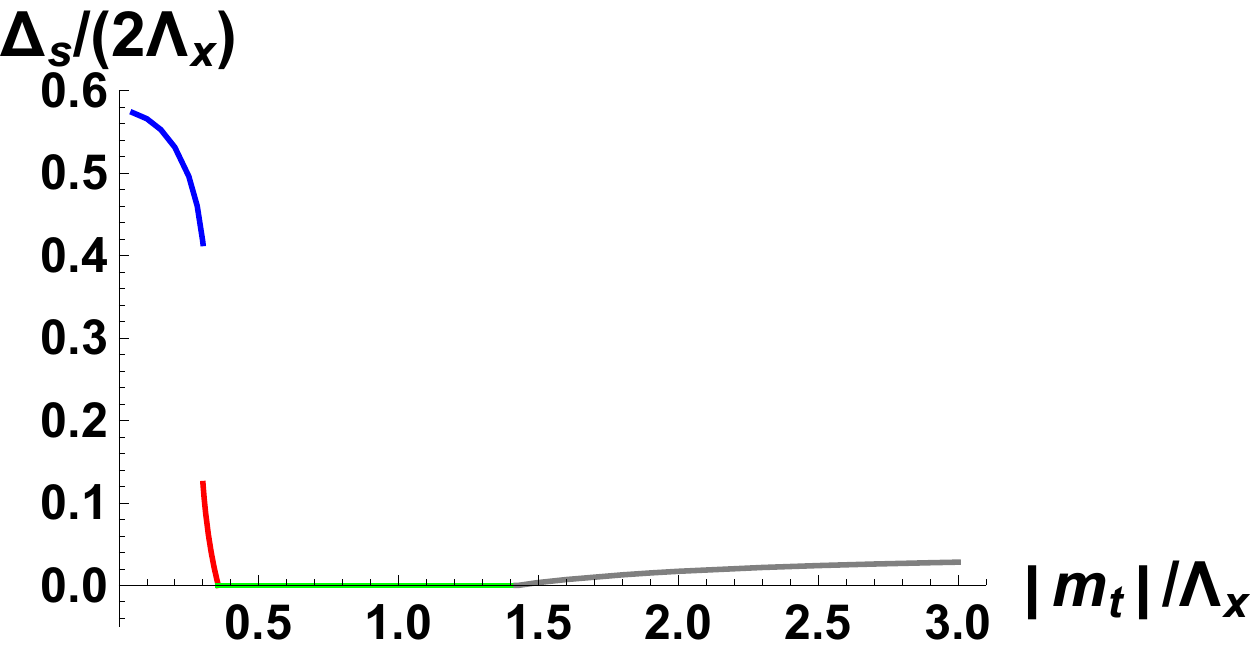}
(d)
\includegraphics[width=0.41\textwidth]{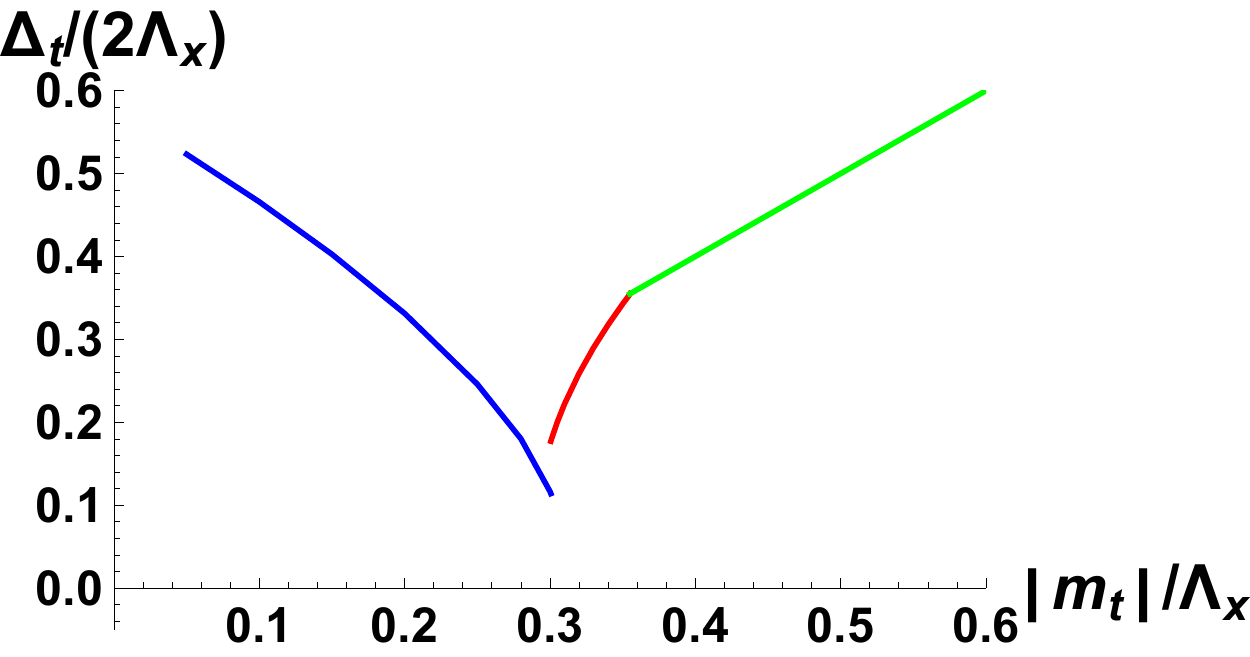}
\caption{(Color online)
(a) and (b)
Cut with fixed $\lambda=\widetilde{\lambda}=4$
from Fig.\ \ref{Fig: schematic phase diagram}(b).
The stable mean-field solutions 
$\Phi^{\,}_{+}$
and
$\Phi^{\,}_{-}$
are presented in panels (a) and (b)
as functions of
$|m^{\,}_{\mathrm{t}}|/\Lambda^{\,}_{x}$, respectively.
The $|m^{\,}_{\mathrm{t}}|/\Lambda^{\,}_{x}$
dependence of the singlet ($\Delta^{\,}_{\mathrm{s}}$)
and the triplet ($\Delta^{\,}_{\mathrm{t}}$)
gaps are plotted in panels (c) and (d)
by making use of Eqs.\
(\ref{eq: singlet and triplet gap function a})
and
(\ref{eq: singlet and triplet gap function b}), respectively.
\label{Fig: details for TRS case example}
         }
\end{center}
\end{figure*}

\subsection{Mean-field phase diagram}
\label{Subec: Mean-field phase diagram}

By combining
Eq.\ (\ref{eq: singlet and triplet anisotropy assumption})
with
Eq.\ (\ref{eq: gap function summary}), we find the singlet and triplet gaps
\begin{subequations}
\label{eq: singlet and triplet gap function}
\begin{align}
\Delta^{\,}_{\mathrm{s}}\:=&\,
2\left|\Phi^{\,}_{-}\right|
=
\left|\phi-\tilde{\phi}\right|,
\label{eq: singlet and triplet gap function a}
\\
\Delta^{\,}_{\mathrm{t}}\:=&\,
2\left||m^{\,}_{\mathrm{t}}|-\left|\Phi^{\,}_{-}\right|\right|
=
2\left||m^{\,}_{\mathrm{t}}|-\frac{1}{2}\left|\phi-\tilde{\phi}\right|\right|,
\label{eq: singlet and triplet gap function b}
\end{align}
\end{subequations}
respectively, given a stable solution $(\Phi^{\,}_{+},\Phi^{\,}_{-})$
to the saddle-point equations
(\ref{eq: the saddle-point equation ms neq 0 befor integral new basis}).
Correspondingly, we enumerate the following four possibilities
\begin{subequations}
\label{eq: MF Ansatz}
\begin{align}
&
\Phi^{\,}_{+}=+\,\Phi^{\,}_{-}=+\phi/2\neq 0,
\qquad
\tilde{\phi}=0,
\label{eq: MF Ansatz (i)}
\\
&
\Phi^{\,}_{+}=-\,\Phi^{\,}_{-}=-\tilde{\phi}/2\neq 0,
\qquad
\phi=0,
\label{eq: MF Ansatz (ii)}
\\
&
\Phi^{\,}_{+}=\phi=\tilde{\phi},
\qquad
\Phi^{\,}_{-}=0,
\label{eq: MF Ansatz (iii)}
\\
&
\Phi^{\,}_{-}=(\phi-\tilde{\phi})/2=\pm m^{\,}_{\mathrm{t}}\neq0.
\label{eq: MF Ansatz (iv)}
\end{align}
\end{subequations}
Case (\ref{eq: MF Ansatz (i)})
is obtained when $\lambda>0$ while $\widetilde{\lambda}=0$.
Case (\ref{eq: MF Ansatz (ii)})
is obtained when $\lambda=0$ while $\widetilde{\lambda}\neq0$.
Case (\ref{eq: MF Ansatz (iii)}) implies that
the singlet gap vanishes,
$\Delta^{\,}_{\mathrm{s}}=0$,
while the triplet gap is solely controlled by the triplet mass,
$\Delta^{\,}_{\mathrm{t}}=|m^{\,}_{\mathrm{t}}|$.
Case (\ref{eq: MF Ansatz (iv)})
implies that the triplet gap vanishes, $\Delta^{\,}_{\mathrm{t}}=0$,
while the singlet gap is solely controlled by the triplet mass,
$\Delta^{\,}_{\mathrm{s}}\neq2|m^{\,}_{\mathrm{t}}|$.

Figure \ref{Fig: schematic phase diagram}
summarizes the numerical search for the stable solutions to
the saddle-point equations
(\ref{eq: the saddle-point equation ms neq 0 befor integral new basis})
in the three-dimensional coupling space
$\lambda\geq0$,
$\widetilde{\lambda}\geq0$,
and $|m^{\,}_{\mathrm{t}}|\geq0$,
holding $v^{\,}_{\mu}$ and $m^{\,}_{\mathrm{s}}$ fixed to the values 
$v^{\,}_{\mu}\equiv1$
and
$m^{\,}_{\mathrm{s}}=0$, respectively.
The terminology ATO for Abelian topological order and NATO
for non-Abelian topological order applies whenever the stable saddle-point
delivers Chern insulating bands with four and one chiral Majorana edge states,
respectively, upon imposing open boundary condition
along the $y$-direction. Which chirality is
to be found on the left ($\mathtt{m}=1$) or right ($\mathtt{m}=n$)
ends of the model defined in Eq.\ (\ref{eq: coupled Majorana theory})
is specified by the combination of letters LR or RL.
Of course, there is no topological order at the mean-field level
as the ground state is non-degenerate when periodic boundary conditions
are imposed. However, we conjecture that the ground state manifolds
in the ATO and NATO phases acquire distinct non-trivial
topological degeneracies when the mean-field approximation is relaxed.
Computing these topological degeneracies is beyond the scope of this paper.

\subsubsection{Phase transitions between ATO and NATO}

There are two wings of yellow-colored surfaces in Fig.\ \ref{Fig: schematic phase diagram}. 
Within the same ``LR-'' or ``RL-'' topologically ordered phases, 
ATO and NATO phases are separated by a yellow-colored surface
on which the triplet gap $\Delta^{\,}_{\mathrm{t}}$ 
defined in Eq.\ (\ref{eq: singlet and triplet gap function b}) vanishes (namely, $|\Phi^{\,}_{-}|=|m^{\,}_{\mathrm{t}}|$).
As a demonstration, we plot in Fig.\ \ref{Fig: details for transition between ATO and NATO} 
the blue curves 
by fixing
$(\lambda,\widetilde{\lambda})=(7,1)$ 
in Fig.\ \ref{Fig: schematic phase diagram}.
In Fig.\ \ref{Fig: details for transition between ATO and NATO}(a)-(b)
We find a continuous dependence on $m^{\,}_{\mathrm{t}}$
of the stable solution $\Phi^{\,}_{+}$ and $\Phi^{\,}_{-}$
to the saddle point equations
(\ref{eq: the saddle-point equation ms neq 0 befor integral new basis}).
It follows from Eq.\ (\ref{eq: singlet and triplet gap function}) 
that the singlet gap $\Delta^{\,}_{\mathrm{s}}$  
and the triplet gap $\Delta^{\,}_{\mathrm{t}}$ 
in Fig.\ \ref{Fig: details for transition between ATO and NATO}(c)-(d)  
are also continuous dependent on $m^{\,}_{\mathrm{t}}$.
Moreover,  
the triplet gap vanishes at $|m^{\,}_{\mathrm{t}}|/\Lambda^{\,}_{x}\approx0.76$ 
that signals a continuous quantum phase transition.

The two yellow wings to the left and right of the quadrant 
$\lambda=\tilde{\lambda}$ in Fig.\ \ref{Fig: schematic phase diagram}
are connected by a stripe (colored in brown) that 
separates the ATO from the NATO phases by a discontinuous quantum phase transition.
As a demonstration,
in the red curves of Fig.\ \ref{Fig: details for transition between ATO and NATO},
we move away from $(\lambda,\widetilde{\lambda})=(4,4)$ 
in Fig.\ \ref{Fig: schematic phase diagram} by   
choosing $(\lambda,\widetilde{\lambda})=4\sqrt{2}(\sin\theta,\cos\theta)$ with $\theta=23\pi/90$.
We present the stable solution $\Phi^{\,}_{+}$ and $\Phi^{\,}_{-}$ as a function of $m^{\,}_{\mathrm{t}}$
in the red curves of Fig.\ \ref{Fig: details for transition between ATO and NATO}(a) and (b), respectively.
There is a discontinuous dependence on $m^{\,}_{\mathrm{t}}$ of the stable solution $\Phi^{\,}_{+}$ and $\Phi^{\,}_{-}$ 
to the saddle point equations 
(\ref{eq: the saddle-point equation ms neq 0 befor integral new basis})
that delivers a discontinuous dependence on $m^{\,}_{\mathrm{t}}$ of 
the singlet gap $\Delta^{\,}_{\mathrm{s}}$ and the triplet gap $\Delta^{\,}_{\mathrm{t}}$
in the red curves of Fig.\ \ref{Fig: details for transition between ATO and NATO}(c) and (d).

\subsubsection{Case $\lambda=\tilde{\lambda}$}

Figure \ref{Fig: schematic phase diagram}(b)
summarizes the numerical search for the stable solutions to
the saddle-point equations
(\ref{eq: the saddle-point equation ms neq 0 befor integral new basis})
in the quadrant
$\lambda=\widetilde{\lambda}\geq0$,
and $|m^{\,}_{\mathrm{t}}|\geq0$,
holding $v^{\,}_{\mu}$ and $m^{\,}_{\mathrm{s}}$ fixed to the values 
$v^{\,}_{\mu}\equiv1$
and
$m^{\,}_{\mathrm{s}}=0$, respectively.
We found three distinct mean-field phases
whose boundaries are shown in
Fig.\ \ref{Fig: schematic phase diagram}(b).
One phase is gapless. 
Two phases are gapful when
periodic boundary conditions are imposed.
The region bounded by the vertical axis and the green curve
supports a stable solution to
the saddle-point equations
(\ref{eq: the saddle-point equation ms neq 0 befor integral new basis})
with $\Phi^{\,}_{+}\neq0$ but $\Phi^{\,}_{-}=0$. Hence, this solution
respects the time-reversal symmetry of the mean-field Hamiltonian.
It follows from Eq.\ (\ref{eq: singlet and triplet gap function}) that
the triplet gap $\Delta^{\,}_{\mathrm{t}}$ is non-vanishing
while the singlet gap $\Delta^{\,}_{\mathrm{s}}$ is vanishing.
The triplet of Majorana are thus gapped, while the singlet of Majorana
is gapless because of a Dirac-like band touching.
The dashed line (colored in brown) in Fig.\ \ref{Fig: schematic phase diagram}(b)
is a line of discontinuous quantum phase transitions 
by which $|\Phi^{\,}_{-}|<|m^{\,}_{\mathrm{t}}|$ above the dashed line,
while $|\Phi^{\,}_{-}|>|m^{\,}_{\mathrm{t}}|$ below the dashed line. 
The discontinuous jump of $|\Phi^{\,}_{-}|$
is evidence for a mean-field discontinuous quantum phase transition.
This discontinuity is mirrored in the discontinuities
of $\Phi^{\,}_{+}$, $\Delta^{\,}_{\mathrm{s}}$, and $\Delta^{\,}_{\mathrm{t}}$
as exemplified in Fig.\ \ref{Fig: details for TRS case example} for $(\lambda,\tilde{\lambda})=(4,4).$

\begin{figure}[t]
\begin{center}
(a)
\includegraphics[width=0.45\textwidth]{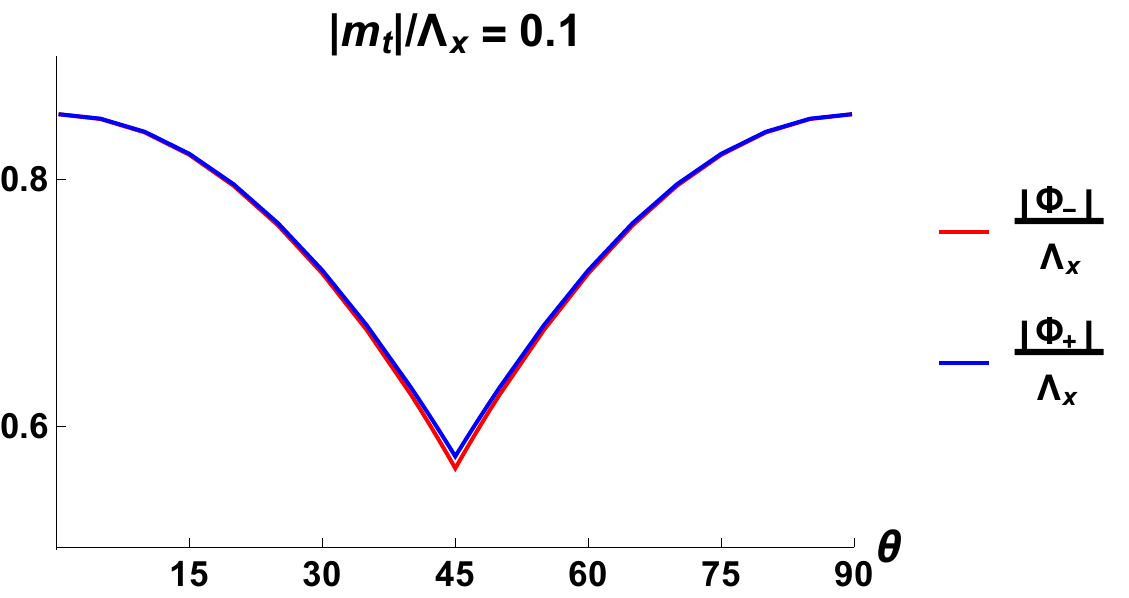}
\\
(b)
\includegraphics[width=0.45\textwidth]{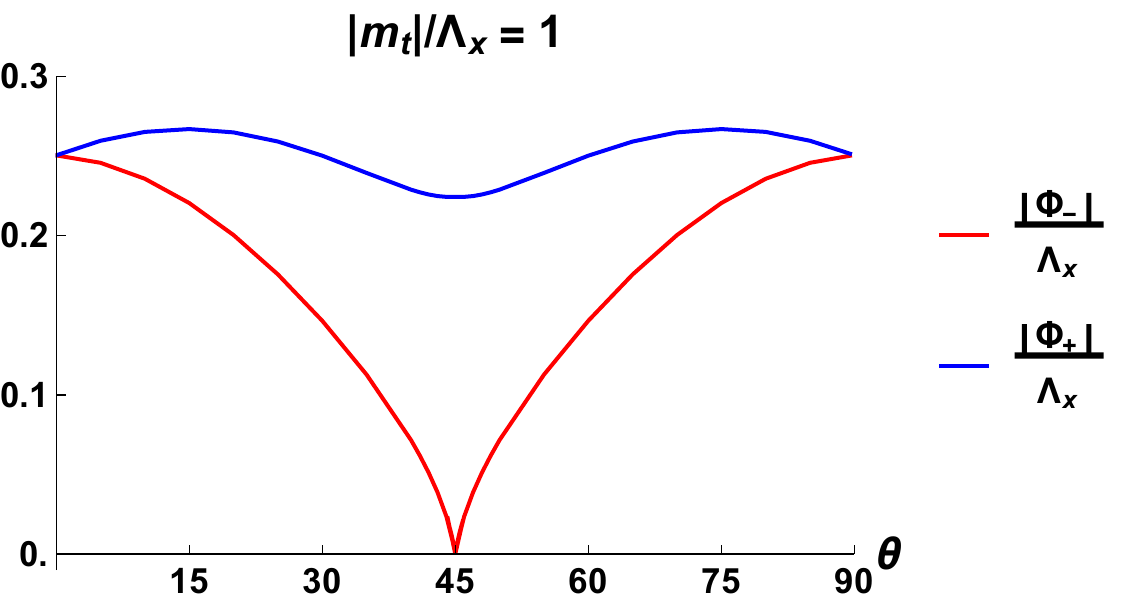}
\\
(c)
\includegraphics[width=0.45\textwidth]{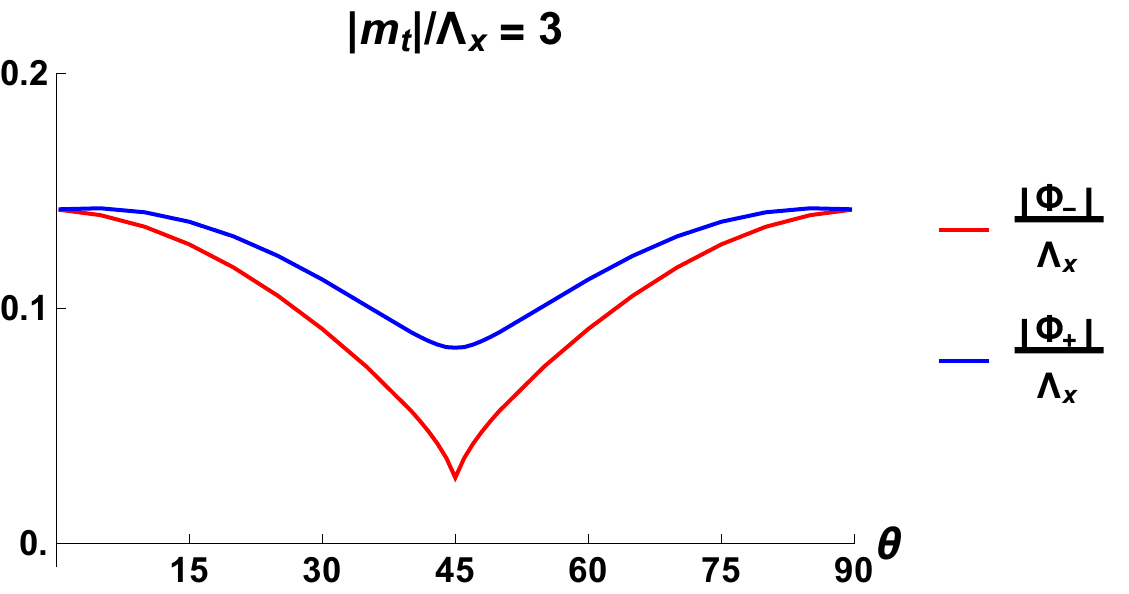}
\caption{(Color online)
The stable mean-field solutions $\Phi^{\,}_{+}$ and $\Phi^{\,}_{-}$ 
as a function of $\theta:=\mathrm{arctan}(\lambda/\widetilde{\lambda})$ and fixing $\lambda^2+\widetilde{\lambda}^2= 32.$
When $\theta=\pi/4,$
there is a discontinuous (respectively, continuous) 
phase transition for panel (a) and (c) [respectively, (b)].
(a)
Case $|m^{\,}_{\mathrm{t}}|/\Lambda^{\,}_{x}=0.1$.
Here, $|\Phi^{\,}_{+}|\neq|\Phi^{\,}_{-}|$ while
$||\Phi^{\,}_{+}|-|\Phi^{\,}_{-}||\ll1$.
(b)
Case $|m^{\,}_{\mathrm{t}}|/\Lambda^{\,}_{x}=1$.
(c)
Case $|m^{\,}_{\mathrm{t}}|/\Lambda^{\,}_{x}=3$.
\label{Fig: order parameter theta different m}
         }
\end{center}
\end{figure}

As a comparison, we plot in Fig.\ \ref{Fig: order parameter theta different m}
the stable mean-field solutions $\Phi^{\,}_{+}$ and $\Phi^{\,}_{-}$ 
as a function of $\theta\:=\mathrm{arctan}(\lambda/\widetilde{\lambda})$ and fixing $\lambda^2+\widetilde{\lambda}^2= 32$
for $|m^{\,}_{\mathrm{t}}|/\Lambda^{\,}_{x}=0.1,1,$ and $3$ in Fig.\ \ref{Fig: order parameter theta different m}(a),(b) and (c), respectively. 
When $\theta=\pi/4,$
there is a discontinuous (respectively, continuous) 
phase transition for panel (a) and (c) [respectively, (b)].
We note that in Fig.\ \ref{Fig: order parameter theta different m}(a), 
the value of $|\Phi^{\,}_{-}|$ is not equal to $|\Phi^{\,}_{+}|$ while
$||\Phi^{\,}_{+}|-|\Phi^{\,}_{-}||\ll1$.

\begin{figure}[t]
\begin{center}
(a)
\includegraphics[width=0.4\textwidth]{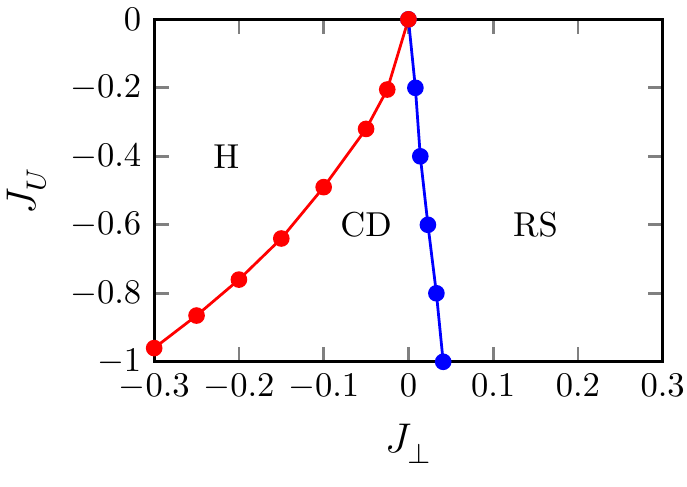}
\qquad\;
\\
(b)
\includegraphics[width=0.21\textwidth]{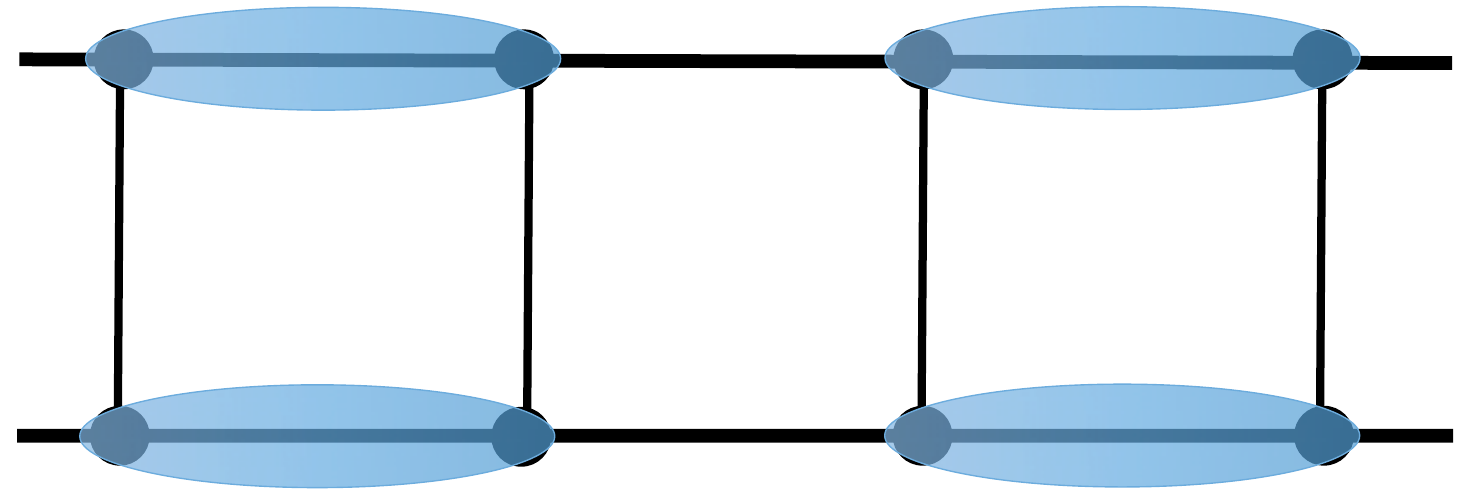}
(c)
\includegraphics[width=0.21\textwidth]{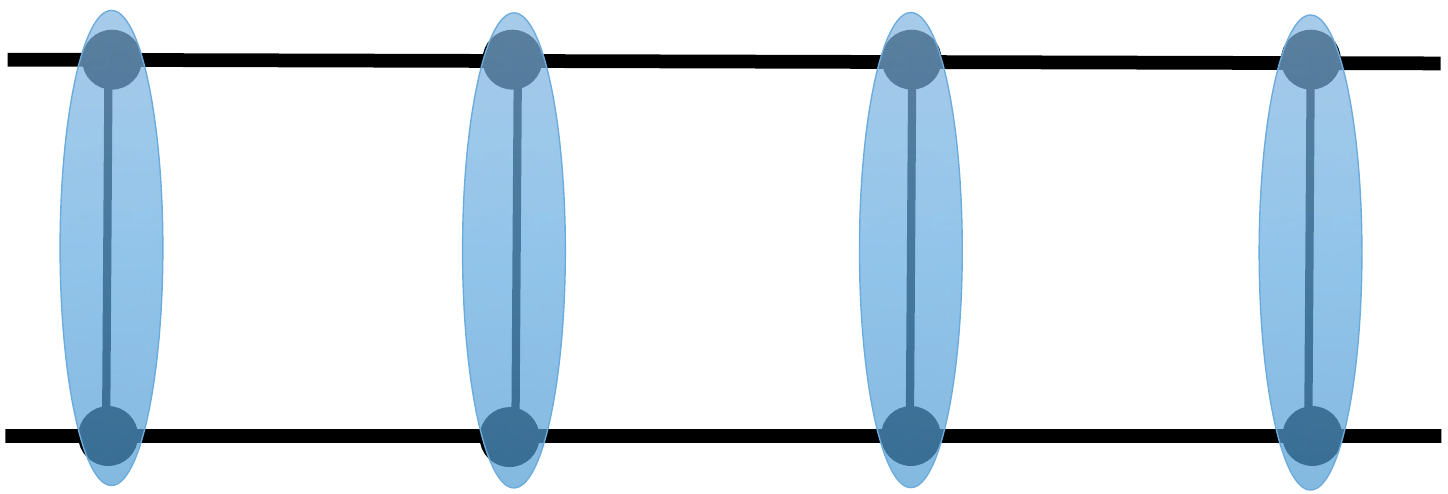}
\caption{(Color online)
(a)
The phase diagram for the ladder model (\ref{eq: def H ladder})
as a function of $J^{\,}_{U}<0$ and $|J^{\,}_{\perp}|\leq0.3.$
The phase boundary between the columnar-dimer (CD) and the rung-singlet (RS)
phases is a continuous phase transition 
in the Ising universality class.
The phase boundary between the Haldane phase (H) 
and the columnar-dimer phase
is a continuous phase transition 
in the $\widehat{su}(2)^{\,}_{2}$ WZNW universality class.
(b)
Classical representation for the CD order.
(c)
Classical representation for the RS order.
\label{Fig: single ladder phase diagram}
         }
\end{center}
\end{figure}

\begin{figure}[t]
\begin{center}
\includegraphics[width=0.5\textwidth]{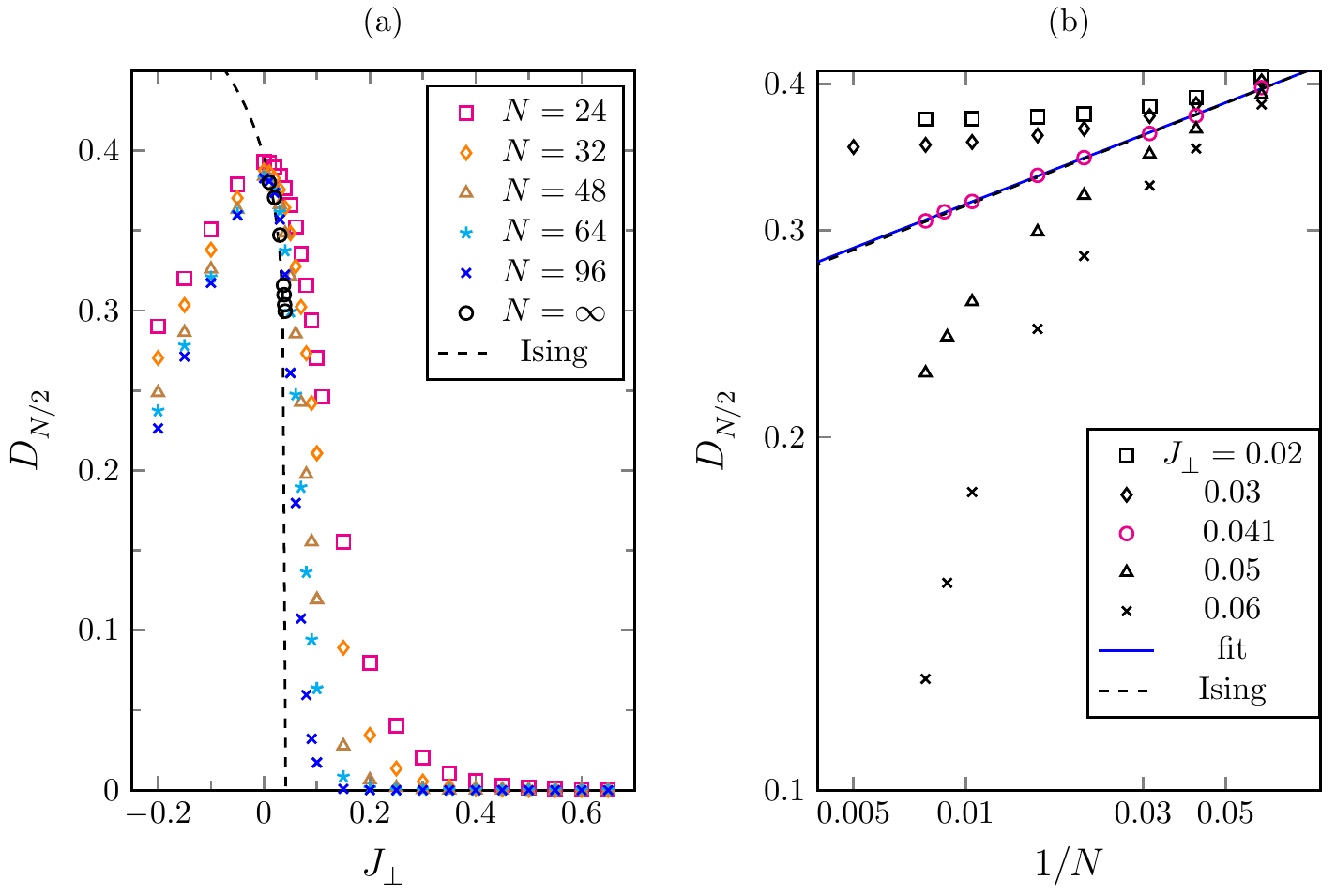}
\includegraphics[width=0.5\textwidth]{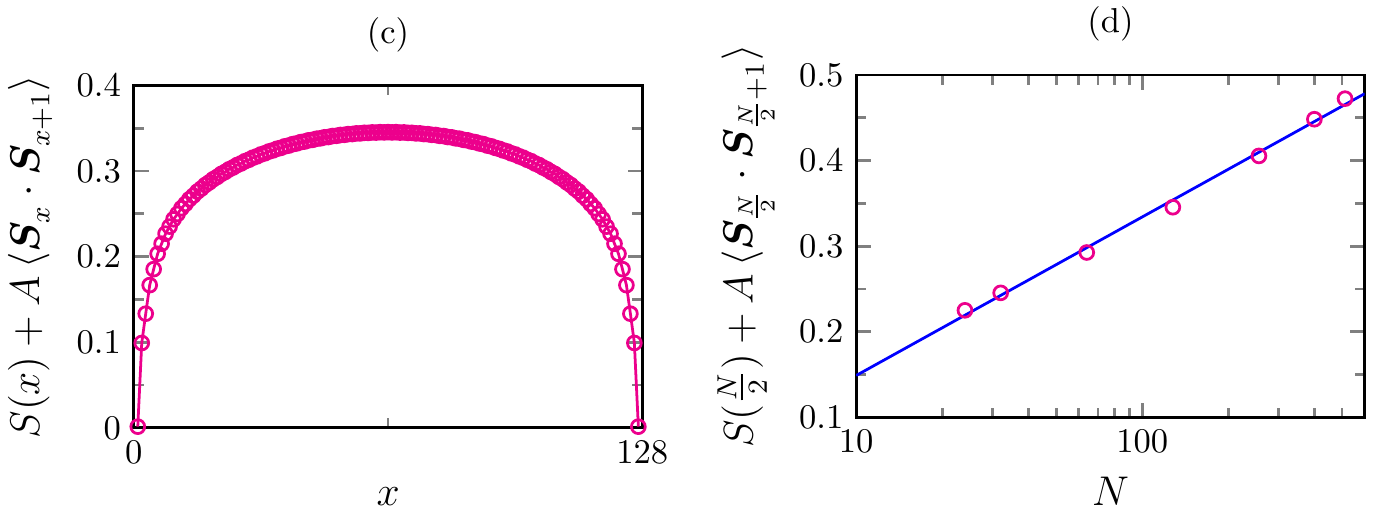}
\caption{(Color online)
(a)
Semi-log plot for the leg-dimer order parameter (\ref{eq: define leg dimer op}) at the center of the ladder,
$D^{\,}_{N/2}$, as a function of $J^{\,}_{\perp}$ for different system sizes while fixing $J^{\,}_{U} = -1$.  
The extrapolation to the thermodynamic limit
is obtained with a second-order polynomial in $1/N$, whereas the dashed
curve is a fit to the Ising scaling law $D^{\,}_{N/2}\propto N^{-1/8}$.
The transition point is at $J^{\,}_{\perp} \approx 0.041$.
(b)
Fixing $J^{\,}_{U} = -1$, this 
log-log plot shows the 
scaling of $D^{\,}_{N/2}$ with $1/N$ for different values of $J^{\,}_{\perp}$ 
in the vicinity of the critical point.
The Ising scaling law fits pretty well the scaling of $D^{\,}_{N/2}$ at 
the transition point $J^{\,}_{\perp} \approx 0.041$.
(c-d)
Fitting the entanglement-entropy from Eq.\ (\ref{eq: entanglement entropy})   
as a function of
$x$ with $N=128$ in pane (c) and
of $N$ with $x=N/2$ in panel (d)
yield $c=0.4692$ and $c=0.4824$, respectively.
\label{Fig: FSS}
         }
\end{center}
\end{figure}

\section{Lattice regularization}
\label{sec: Implications for lattice models}

We are going to show that the 
one-dimensional lattice model
(\ref{eq: def H ladder})
regularizes the $(1+1)$-dimensional quantum field theory with
the Hamiltonian density obtained by adding
Eq.\ (\ref{eq: coupled Majorana theory kinetic})
to
Eq.\ (\ref{eq: coupled Majorana theory intraladder})
with $n=1$.
This will be achieved using
the density matrix renormalization group (DMRG)
\cite{white92,white93}
to match quantum criticality in
the quantum field theory with that in the lattice model.

We will then couple a one-dimensional array of spin-1/2 ladders
of the form (\ref{eq: def H ladder})
as is done in Hamiltonian
(\ref{eq: def couplings between consecutive ladders})
and argue that this two-dimensional lattice model
regularizes the Hamiltonian density
(\ref{eq: coupled Majorana theory}).

\subsection{Numerical study of a two-leg ladder}
\label{Numerical study of a two-leg ladder}

Following Ref.\ \onlinecite{Chen17},
we define a spin-1/2 ladder by the Hamiltonian
\begin{align}
\widehat{H}^{\,}_{\mathrm{ladder}}\:=
&\,
\sum_{i=1}^{N-1}
J^{\,}_{1}\,
\widehat{\bm{S}}^{\,}_{i}
\cdot
\widehat{\bm{S}}^{\,}_{i+1}
+
\sum_{i^{\prime}=1}^{N-1}
J^{\,}_{1}\,
\widehat{\bm{S}}^{\prime}_{i^{\prime}}
\cdot
\widehat{\bm{S}}^{\prime}_{i^{\prime}+1}
\nonumber\\
&\,
+
\sum_{i=1}^{N}
J^{\,}_{\perp}\,
\widehat{\bm{S}}^{\,}_{i}
\cdot
\widehat{\bm{S}}^{\prime}_{i}
\nonumber\\
&\,
+
\sum_{i=1}^{N-1}
J^{\,}_{\times}
\left(
\widehat{\bm{S}}^{\,}_{i}
\cdot
\widehat{\bm{S}}^{\prime}_{i+1}
\,+\,
\widehat{\bm{S}}^{\,}_{i+1}
\cdot
\widehat{\bm{S}}^{\prime}_{i}
\right)
\nonumber\\
&\,
+
\sum_{i=1}^{N-1}
J^{\,}_{U}\,
\left(
\widehat{\bm{S}}^{\,}_{i}
\cdot
\widehat{\bm{S}}^{\,}_{i+1}
\right)
\left(
\widehat{\bm{S}}^{\prime}_{i}
\cdot
\widehat{\bm{S}}^{\prime}_{i+1}
\right).
\label{eq: def H ladder}
\end{align}
Here, $\widehat{\bm{S}}^{\,}_{i}$ and $\widehat{\bm{S}}^{\prime}_{i'}$
are spin-1/2 operators localized on the sites of
the first and second legs of the ladder, respectively.
There are three independent couplings obeying
$J^{\,}_{1}>0$
and $J^{\,}_{\perp},J^{\,}_{\times},J^{\,}_{U}\in\mathbb{R}$
with the condition
$J^{\,}_{\times}\equiv-J^{\,}_{\perp}/2$.
References \cite{Shelton96,Nersesyan97,Chen17}
have shown that, at the level of bosonization,
the low-energy limit of the ladder (\ref{eq: def H ladder})
is the single copy ($n=1$) of
the non-interacting massive Majorana field theory 
defined by adding the Hamiltonian densities
(\ref{eq: coupled Majorana theory kinetic})
and (\ref{eq: coupled Majorana theory intraladder})
with the mass terms $m^{\,}_{\mathrm{s}}$ and $m^{\,}_{\mathrm{t}}$
related to the microscopic couplings in Eq.\ (\ref{eq: def H ladder})
by
\begin{align}
m^{\,}_{\mathrm{s}}=\frac{-1}{2\pi}
\left(12J^{\,}_{\perp}+J^{\,}_{U}\right),
\qquad
m^{\,}_{\mathrm{t}}=\frac{1}{2\pi}
\left(4J^{\,}_{\perp}-J^{\,}_{U}\right).
\label{eq: ladder couplings and mass}
\end{align}
Bosonization thus predicts the existence for
the spin-1/2 ladder (\ref{eq: def H ladder})
of a quantum critical point
in the Ising universality class
for which $m^{\,}_{\mathrm{s}}=0$
as the dimensionless ratio
$J^{\,}_{U}/J^{\,}_{\perp}$
smoothly crosses the critical value
$\left(J^{\,}_{U}/J^{\,}_{\perp}\right)^{\,}_{\mathrm{c}}\approx-1/12$.
We are going to use the technique of
the density matrix renormalization group (DMRG)
\cite{white92,white93}
to verify this prediction.
We fix the units of energy by setting $J^{\,}_{1}=1$,
bound from above the bond dimension in the DMRG by 1500,
and impose open boundary condition.

The phase diagram as a function of $J^{\,}_{U}<0$
and $|J^{\,}_{\perp}|\leq0.3$ is shown in Fig.\
\ref{Fig: single ladder phase diagram}(a).
Here, CD and RS stand for columnar-dimer and rung-singlet,
respectively. A classical representation for the CD and
the RS phases is obtained by coloring
nearest-neighbor bonds as shown
in Figs.\ \ref{Fig: single ladder phase diagram}(b)
and \ref{Fig: single ladder phase diagram}(c).
The acronym H stands for the Haldane phase of the antiferromagnetic
quantum spin-1 Heisenberg chain \cite{Haldane83a,Haldane83b}.
The Haldane phase is obtained when
$J^{\,}_{\perp}$ is ferromagnetic ($J^{\,}_{\perp}<0$)
and $|J^{\,}_{U}|$ is not too large. Increasing
$|J^{\,}_{U}|$ weakens the Haldane phase until it gives way
to the CD phase. Destroying the CD phase is achieved by
changing the sign of $J^{\,}_{\perp}$ holding $|J^{\,}_{U}|$
fixed.

The phase boundary between the CD phase and
the RS phase is a continuous phase transition belonging to the
two-dimensional Ising universality class.
The numerical evidence for this Ising transition is
supported by the finite-size scaling of the leg-dimer order parameter
\cite{Lavarelo11,Huang17}
\begin{align}
D^{\,}_{i}\:=
\left\langle
\widehat{\bm{S}}^{\,}_{i}\cdot
\left(
\widehat{\bm{S}}^{\,}_{i+1}
-\widehat{\bm{S}}^{\,}_{i-1}
\right)
\right\rangle,
\quad i =1,\cdots,N-1,
\label{eq: define leg dimer op}
\end{align}
combined with an estimate of the central charge
from the scaling of the entanglement entropy.

In Figs.\ \ref{Fig: FSS}(a) and \ref{Fig: FSS}(b), we fix $J^{\,}_{U}=-1$.
We then calculate $D^{\,}_{N/2}$ for various value of $J^{\,}_{\perp}$.
We find an Ising critical point at $J^{\,}_{\perp}\approx0.041$
for which the Ising scaling law for the order parameter,
$D^{\,}_{N/2}\propto N^{-1/8}$,
provides an excellent fit.

Another piece of evidence to support the Ising transition is provided
by the scaling form of the bipartite von~Neumann entanglement entropy
under open boundary
condition.
\cite{Calabrese04,Calabrese09,Laflorencie06,Affleck09,Cardy10,Lavarelo11}
It is given by
\begin{equation}
S(x,N)=
\frac{c}{6}
\ln
\left(
\frac{N+1}{\pi}\sin\frac{\pi\,x}{N+1}
\right)
+
A
\langle 
\widehat{\bm{S}}^{\,}_{x}\cdot\widehat{\bm{S}}^{\,}_{x+1}
\rangle
+
B.
\label{eq: entanglement entropy} 
\end{equation}
Here, $x$ is the position of the rung at which we partition the ladder
into left and right ``worlds'',
$c$ is the (to be determined) central charge,
and $A,B$ are non-universal constants.
In Figs.\ \ref{Fig: FSS}(c) and \ref{Fig: FSS}(d),
we fix $J^{\,}_{U}=-1$ and $J^{\,}_{\perp}=0.041$.
In Fig.\ \ref{Fig: FSS}(c), we vary $x$ keeping $N$ fixed.
In Fig.\ \ref{Fig: FSS}(d), we fix $x=N/2$ and vary $N$. 
Both calculations are consistent
with an Ising transition for which the exact central charge
$c=1/2$.

The phase boundary between the H phase and
the RS phase in Fig.\
\ref{Fig: single ladder phase diagram}(a)
is predicted within the bosonization framework
to be a continuous phase transition belonging to the (1+1)-dimensional
$\widehat{su}(2)^{\,}_{2}$ Wess-Zumino-Novikov-Witten (WZNW)
universality class.  The central charge is $3/2$ and the critical
exponent for the scaling of the order parameter is $3/8$. We have
obtained DMRG evidence for such a transition in the same
way as was done for the Ising transition. As this transition is
not the focus of this paper,
we will not present these numerical results.

We conclude this section by observing that 
the spin-ladder model defined in Eq.\ (\ref{eq: def H ladder})
with $J^{\,}_{\times}\equiv0$
was recently studied in Ref.\ \onlinecite{Robinson18}.
Reference\ \onlinecite{Robinson18}
derives a phase diagram similar to that shown
in Fig.\ \ref{Fig: single ladder phase diagram}(a).
The only differences are the slopes of the phase boundaries.
These differences can be understood from the fact that the
phase boundaries of the spin-1/2 ladder (\ref{eq: def H ladder})
are determined by the zeros of the masses of the Majorana fields
(\ref{eq: ladder couplings and mass}).
Choosing different intra-ladder couplings changes the relation
(\ref{eq: ladder couplings and mass})
between the masses of the Majorana fields
and the microscopic couplings.
This change affects the slopes of the phase boundaries
in the microscopic model.
We opted to introduce a non-vanishing coupling
$J^{\,}_{\times}\equiv-J^{\,}_{\perp}/2$
in Hamiltonian (\ref{eq: def H ladder})
in order to suppress all the bare couplings for all marginally relevant
perturbations to the $\widehat{su}(2)_{1}\oplus\widehat{su}(2)_{1}$
WZWN critical point [i.e., all couplings except $J^{\,}_{1}$ set to zero
in Eq.\ (\ref{eq: def H ladder})].%
~\cite{Shelton96,Chen17}

\subsection{Model of coupled spin-1/2 two-leg ladders}
\label{subsec: Model of coupled spin-1/2 two-leg ladders}

We take $n$-copies labeled by the index $\mathtt{m}=1,\cdots,n$
of the spin-1/2 ladder (\ref{eq: def H ladder}).
We couple this array of spin-1/2 ladders with the inter-ladder interaction%
~\cite{Chen17}
\begin{subequations}
\label{eq: def couplings between consecutive ladders}
\begin{equation}
\widehat{H}^{\,}_{\text{inter-ladder}}\:=
\widehat{H}^{\,}_{\triangle}+\widehat{H}^{\prime}_{\triangle}
+
\widehat{H}^{\,}_{\square}+\widehat{H}^{\prime}_{\square},
\end{equation}  
where
\begin{align}
\widehat{H}^{\,}_{\triangle}\:=&\,
\frac{J^{\,}_{\chi}}{2}\,
\sum_{i=1}^{N}
\sum_{\mathtt{m}=1}^{n-1}
\Big[
\widehat{\bm{S}}^{\,}_{i,\mathtt{m}+1}
\cdot
\left(
\widehat{\bm{S}}^{\,}_{i+1,\mathtt{m}}
\wedge
\widehat{\bm{S}}^{\,}_{i,\mathtt{m}}
\right)
\nonumber\\
&\,
+
\widehat{\bm{S}}^{\,}_{i+1,\mathtt{m}}
\cdot
\left(
\widehat{\bm{S}}^{\,}_{i,\mathtt{m}+1}
\wedge
\widehat{\bm{S}}^{\,}_{i+1,\mathtt{m}+1}
\right)
\Big]
\label{eq: def couplings between consecutive ladders a}
\end{align}
and
\begin{align}
\widehat{H}^{\,}_{\square}\:=&\,
J^{\,}_{\vee}\,
\sum_{i=1}^{N}
\sum_{\mathtt{m}=1}^{n-1}
\Big(
\widehat{\bm{S}}^{\,}_{i,\mathtt{m}}
\cdot
\widehat{\bm{S}}^{\,}_{i,\mathtt{m}+1}
\nonumber\\
&\,
+
\frac{1}{2}
\widehat{\bm{S}}^{\,}_{i,\mathtt{m}+1}
\cdot
\widehat{\bm{S}}^{\,}_{i+1,\mathtt{m}}
+
\frac{1}{2}
\widehat{\bm{S}}^{\,}_{i,\mathtt{m}}
\cdot
\widehat{\bm{S}}^{\,}_{i+1,\mathtt{m}+1}
\Big),
\label{eq: def couplings between consecutive ladders b}
\end{align}
\end{subequations} 
with $\widehat{H}^{\prime}_{\triangle}$ and
$\widehat{H}^{\prime}_{\square}$ deduced from
$\widehat{H}^{\,}_{\triangle}$ and $\widehat{H}^{\,}_{\square}$ by the
substitution
$\widehat{\bm{S}}^{\,}_{i,\mathtt{m}}\to\widehat{\bm{S}}^{\prime}_{i,\mathtt{m}}$.
The low energy limit of Hamiltonian
$\widehat{H}^{\,}_{\text{inter-ladder}}$
was obtained using bosonization in
Ref.\ \onlinecite{Huang17,Chen17}
(see also Ref.\ \onlinecite{Gorohovsky15}).
Aside from a renormalization of the velocities entering
the quadratic Hamiltonian density
(\ref{eq: coupled Majorana theory kinetic}),
it produces, as was shown in Ref.\ \onlinecite{Tsvelik90},
the quartic Majorana interaction
(\ref{eq: coupled Majorana theory interladder})
with the couplings $\lambda$ and $\widetilde{\lambda}$
related to the microscopic couplings entering
Eq.\ (\ref{eq: def couplings between consecutive ladders a})
by%
~\cite{Huang17,Chen17}
\begin{subequations}
\label{eq: continuum limit of the coupled ladders b}
\begin{align}
\lambda=2\mathfrak{a}\left[(J^{\,}_{\chi}/\pi)+2J^{\,}_{\vee}\right],
\quad
\widetilde{\lambda}=2\mathfrak{a}\left[-(J^{\,}_{\chi}/\pi)+2J^{\,}_{\vee}\right].
\end{align}
We will use shortly the reciprocal relation
\begin{align}
J^{\,}_{\vee}=\frac{1}{8\mathfrak{a}}\left(\lambda+\widetilde{\lambda}\right),
\quad
J^{\,}_{\chi}=\frac{\pi}{4\mathfrak{a}}\left(\lambda-\widetilde{\lambda}\right).
\end{align}
\end{subequations}

The two-dimensional spin-1/2 model is then defined by
\begin{equation}
\widehat{H}\:=
\widehat{H}^{\mathrm{array}}_{\mathrm{ladder}}
+
\widehat{H}^{\,}_{\text{inter-ladder}},
\label{eq: def 2dim spin-1/2 lattice model}
\end{equation}
where $\widehat{H}^{\mathrm{array}}_{\mathrm{ladder}}$ is simply
the sum of $n$ copies of the spin-1/2 ladder
(\ref{eq: def H ladder}).

\subsection{Implications}

We are now ready to deduce
from the mean-field phase diagram Fig.\ \ref{Fig: schematic phase diagram}
of the quantum field theory (\ref{eq: coupled Majorana theory})
the following predictions for the
two-dimensional array of coupled spin-1/2 ladders (\ref{eq: def 2dim spin-1/2 lattice model}).

First, fixing $m^{\,}_{\mathrm{s}}=0$ implies the linear condition 
[c.f. Eq.\ (\ref{eq: ladder couplings and mass})]
\begin{align}
J^{\,}_{\perp}
\propto
-J^{\,}_{U}.
\end{align}
It then follows that $m^{\,}_{\mathrm{t}}$ is only controlled by one parameter,
namely
\begin{align}
|m^{\,}_{\mathrm{t}}|\propto
|J^{\,}_{\perp}|\propto
|J^{\,}_{U}|.
\end{align}

Second, fixing $\lambda=\widetilde{\lambda}$
implies $J^{\,}_{\chi}\equiv0$, i.e., the
three-spin interaction that breaks explicitly time-reversal symmetry must
vanish. We then deduce from the quantum field theory
(\ref{eq: coupled Majorana theory})
that the two-dimensional spin-1/2 lattice model
(\ref{eq: def 2dim spin-1/2 lattice model})
could support three phases, of which two are gapped and break
spontaneously the time-reversal symmetry while one is gapless
and time-reversal symmetric. There is an important a caveat here, namely
that we have neglected perturbations, whose bare couplings are 
very small (e.g., generated by quantum corrections) but relevant at the
$\bigoplus_{\mathtt{m}}[\widehat{su}(2)_{1}\oplus\widehat{su}(2)^{\,}_{1}]$
WZWN critical point, that would stabilize collinear long-ranged ordered phase
or dimer phases.\cite{Starykh04,Starykh07}
If we ignore this possibility, a too small or too large
$|m^{\,}_{\mathrm{t}}|\propto|J^{\,}_{\perp}|\propto|J^{\,}_{U}|$
could then stabilize a topologically ordered spin-liquid phase, 
whereas intermediate values of
$|m^{\,}_{\mathrm{t}}|\propto|J^{\,}_{\perp}|\propto|J^{\,}_{U}|$
with $\lambda=\widetilde{\lambda}\propto~J^{\,}_{\vee}>0$
not too large (say, $\lambda\lesssim8$)
could stabilize a gapless spin-liquid phase with a Dirac point.
The mean-field transition through the time-reversal-symmetric
quadrant $\lambda=\widetilde{\lambda}$ from
the region with $\lambda<\widetilde{\lambda}$
to the region with $\lambda>\widetilde{\lambda}$
is continuous (discontinuous) if it goes through the gapless
(one of the gapped) phase.

\section{Summary}
\label{sec: Summary}

We have studied a strongly interacting quantum field theory (QFT) 
describing a two-dimensional array of wires containing four (a singlet and a triplet) 
massive Majorana fields in $(1+1)$-dimensional spacetime. 
This QFT is a continuum limit of a
two-dimensional lattice model of spins S=1/2 interacting via SU(2)
symmetric two- three- and four spin interactions.  In the continuum
limit these interactions give rise to two Majorana masses and to
competing quartic Majorana interactions (with couplings $\lambda$ and
$\widetilde{\lambda}$) that are interchanged under time reversal.  The
case $\lambda\neq0$, $\widetilde{\lambda}=0$ when the time reversal is
explicitly broken was studied by us before \cite{Chen17}.  Here,
we have considered the limit $\lambda=\widetilde{\lambda}$ and
established the conditions under which time-reversal symmetry is
broken spontaneously.

At the mean-field level on the time-reversal-symmetric plane
$\lambda=\widetilde{\lambda}$.  we have found three competing phases.
There are two gapped phases that break spontaneously the time-reversal
symmetry, They are gapped in the bulk, and support chiral Majorana
edge modes carrying the chiral central charges 2 and 1/2,
respectively.  One phase is conjectured to signal an Abelian
topological order (ATO), the other is conjectured to signal a
non-Abelian topological order (NATO), if the mean-field approximation
is relaxed. This pair of mean-field gapped phases is separated by a
line of points at which a discontinuous phase transition takes
place. However, we have also found a time-reversal-symmetric
mean-field phase that supports a branch of mean-field Majorana modes
with a gapless Dirac spectrum. This phase is bounded by a line of
continuous phase transitions separating it from the mean-field
snapshot of the NATO phase.

We remark that although we have assumed that the singlet mass
$m^{\,}_{\mathrm{s}}$ is vanishing in our mean-field analysis and
treated the triplet mass $m^{\,}_{\mathrm{t}}$ as a tunable parameter,
we could equally well have reversed the roles of the singlet and
triplet masses. If so, we can simply exchange the role played by the
triplet and the singlet Majorana modes.  The resulting mean-field
phase diagram would contain again the mean-field snapshots of an
Abelian phase and of a non-Abelian phase.  The Abelian phase is the
same Abelian phase as in the present study.  The non-Abelian phase
would be different, however, as its chiral edge modes would carry a
chiral central charge of $3/2$.  A non-Abelian topologically ordered
phase with chiral edge states endowed with the central charge $3/2$ is
a cousin to the Moore-Read state for the fractional quantum Hall
effect \cite{Moore91}.  One also finds such a non-Abelian
topologically ordered phase for certain spin-$1$ Heisenberg models on
the square lattice \cite{Chen18}.

\begin{acknowledgments}
The DMRG calculations were performed 
using the ITensor library%
~\footnote{
ITensor library
(version 2.1.1)
\href{http://itensor.org}{http://itensor.org}
          }
on the Euler cluster at ETH Z\"urich, Switzerland.
J.-H.C.\ was supported by the Swiss National Science Foundation (SNSF) 
under Grant No.\ 2000021 153648. 
C.C.\ was supported by the U.S.\ Department of Energy (DOE), 
Division of Condensed Matter Physics and Materials Science, 
under Contract No.\ DE-FG02-06ER46316. 
A.M.T.\ was supported by the U.S.\
Department of Energy, Office of Basic Energy Sciences,
under Contract No.\ DE-SC0012704. 
\end{acknowledgments}

\bibliography{biblioDraft}

\end{document}